\documentclass[times,final]{elsarticle}
\usepackage{jcomp}
\usepackage{framed,multirow}

\usepackage{amssymb}
\usepackage{latexsym}

\usepackage{hyperref}
\usepackage{subcaption}
\usepackage{float}
\usepackage{amsmath}
\usepackage{url}
\usepackage{xcolor}
\usepackage{natbib}
\definecolor{newcolor}{rgb}{.8,.349,.1}

\journal{Journal of Computational Physics}

\begin{document}

\verso{S. Naskar \textit{et al.}}

\begin{frontmatter}

\title{A generalized curvilinear solver for spherical shell Rayleigh-B\'enard convection}%\tnoteref{tnote1}}%

\author[1]{Souvik Naskar} %\snm{A}%
\author[2]{Karu Chongsiripinyo}
%\fnref{fn1}}
%\fntext[fn1]{This is author footnote for second author.}  
\author[1]{Anikesh Pal\corref{cor1}}
\ead{pala@iitk.ac.in}
\cortext[cor1]{Corresponding author: 
  Tel;  
  }
\author[1]{Akshay Jananan}  

\address[1]{Department of Mechanical Engineering, Indian Institute of Technology, Kanpur 208016, India}
\address[2]{Department of Mechanical Engineering, Chulalongkorn University, Bangkok, Thailand}

\received{}
\finalform{}
\accepted{}
\availableonline{}
\communicated{}

\begin{abstract}
%%%

A three-dimensional finite-difference solver has been developed and implemented for Boussinesq convection in a spherical shell. The solver transforms any complex curvilinear domain into an equivalent Cartesian domain using Jacobi transformation and solves the governing equations in the latter. This feature enables the solver to account for the effects of the non-spherical shape of the convective regions of planets and stars. Apart from parallelization using MPI, implicit treatment of the viscous terms using a pipeline alternating direction implicit scheme and HYPRE multigrid accelerator for pressure correction makes the solver efficient for high-fidelity direct numerical simulations. We have performed simulations of Rayleigh-B\'enard convection at three Rayleigh numbers $Ra=10^{5}, 10^{7}$ and $10^{8}$ while keeping the Prandtl number fixed at unity ($Pr=1$).  The average radial temperature profile and the Nusselt number match very well, both qualitatively and quantitatively, with the existing literature. Closure of the turbulent kinetic energy budget, apart from the relative magnitude of the grid spacing compared to the local Kolmogorov scales, assures sufficient spatial resolution. %Grid convergence and scaling of computational speed with the number of processors have also been reported.
%%%%
\end{abstract}

%\begin{keyword}
%% Keywords
%\KWD Keyword1\sep Keyword2\sep Keyword3
%\end{keyword}
\end{frontmatter}

%\linenumbers
\section{Introduction}\label{sec:intro}

\begin{figure}
\centering
\includegraphics[scale=.5]{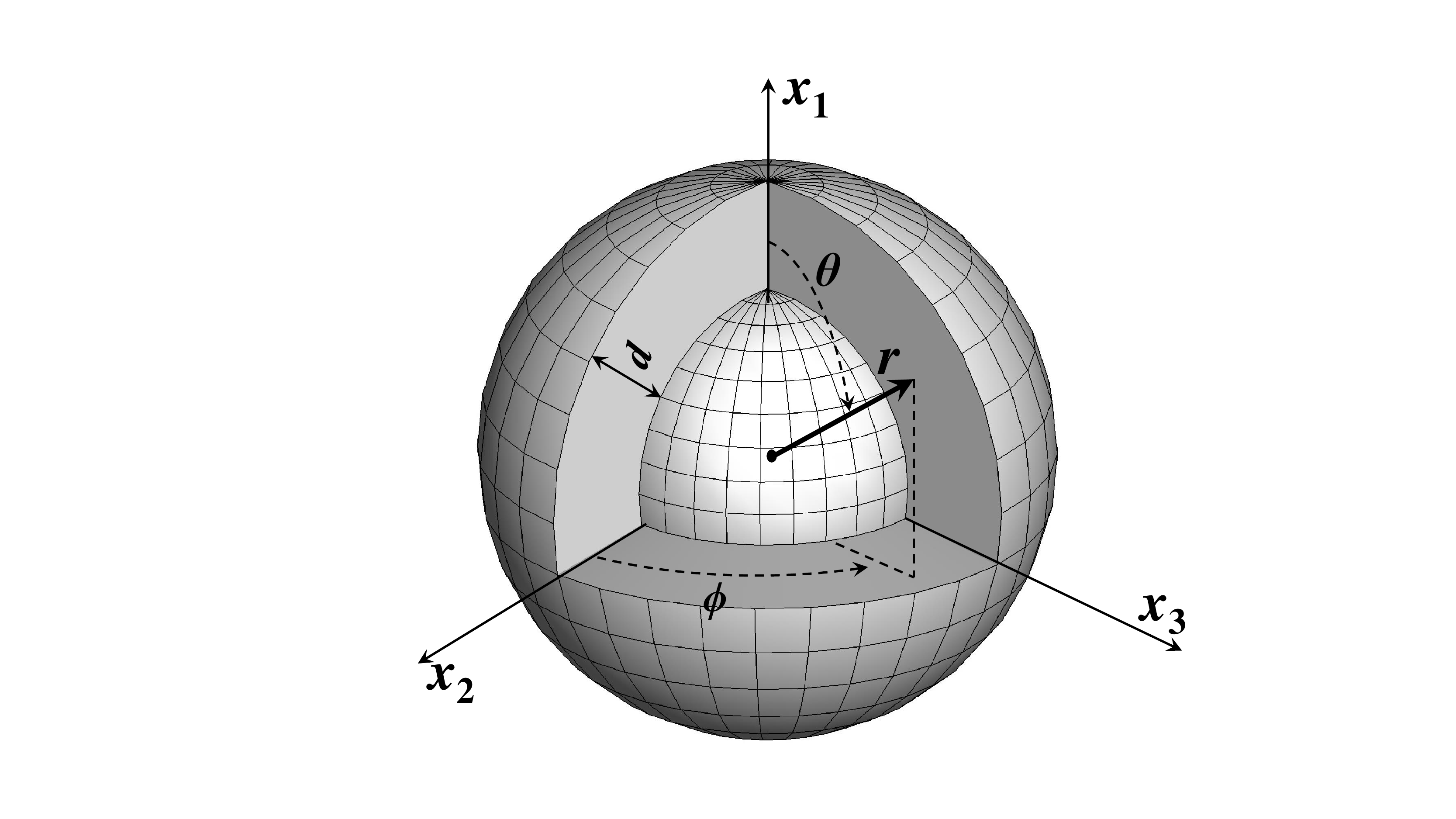}
\caption{Spherical shell geometry.}
\label{fig:spherical domain}
\end{figure}

%%Structure of intro
% Basics and motivations for plane layer RBC
Turbulent thermal convection is ubiquitous in nature as a primary driving mechanism for atmospheric and oceanic circulations \citep{hartmann2001}. Such convective motions in Earth's outer core or in the solar convective zone, for example,  provide energy to sustain global-scale magnetic fields in planets and stars\citep{roberts_2013,rudiger_2006}. Such flow phenomena are further enriched due to the presence of global rotation, external or self-generated magnetic fields, chemical reactions, phase change, the porosity of the medium, and particle suspension \citep{chilla2012}.  Furthermore, the design of heat exchangers, cooling systems for electronics, and indoor air circulation systems requires a fundamental understanding of thermal convection \citep{incropera_1988,incropera_1996}. Rayleigh-B\'enard convection (RBC) is a simple model of thermal convection, where a fluid layer between two parallel plates is heated from below and cooled from above. Such a plane layer geometry can be considered, for example, as a local approximation of the tangent cylinder region  of Earth's outer core, which is situated between the top and bottom surfaces of the solid inner core and extending towards the north and south poles, respectively, up to the core-mantle boundary.  Simulations in the plane layer geometry can reproduce the basic force balance and heat transfer behavior that can be validated from well-designed laboratory experiments. Therefore, this flow configuration has been extensively studied, with the individual or combined effect of global rotation and magnetic fields \citep{naskar2022direct,naskar_pal_2022} to model various geophysical and astrophysical turbulent flows \citep{ahlers2009}.\\  

% Basics and motivations for spherical shell RBC
In the geophysical and astrophysical context, however,  a spherical shell geometry is more pertinent to modeling planetary cores or stellar convective zones. The most extensive body of literature in this geometry focuses on "geodynamo" simulations that attempt to model convection in Earth's outer core convection and the associated geomagnetic field originating from it \citep{jones_2011}. Mantle convection \citep{wolstencroft2009}, rapidly rotating convection \citep{gastine_2016,aurnou_2015}, RBC without rotation and magnetic field \citep{Gastine2015,mound_2017,long_2020b}, deep convection in gas giants\cite{yadav_2020a,yadav2020b}, and solar convection \citep{korre_2021} are among the other prolific areas of research where spherical shell models are implemented. The superiority of these models lies in their capability to model many essential dynamical features of planetary atmospheres, such as thermal winds, strong shear layers, magnetic buoyancy, meridional circulations, and large-scale flows. They can also incorporate important geometric constraints, such as tangent cylinders and curvature effects near the boundaries, whose combined or individual influence can not be accounted for in a local Cartesian plane layer configuration \citep{rincon_2019}. \\

% A comparative discussion of the two geometries
The local plane layer and the global spherical shell simulations differ primarily in the direction of gravity, which is generally kept vertically downwards in the local Cartesian models. In contrast, the direction is radially inwards in global spherical shell models. Additionally, rotating convection in spherical shells exhibits distinct scales in the radial, axial, and azimuthal directions \citep{dormy_2004}, whereas, for the local Cartesian model, we need to consider only two spatial scales: the horizontal scale of convection and the vertical scale over which convection occurs. For both geometries, the governing non-dimensional parameters are the Rayleigh numbers ($Ra$), which is a non-dimensional measure of the thermal forcing, and the Prandtl number ($Pr$), representing the viscous to thermal diffusivity ratio. Apart from this, the flow properties may also depend on the aspect ratio $\Gamma=W/H$ (where W and H are the horizontal and vertical extents of the domain) and the radius ratio $\Gamma=r_i/r_o$ in-plane layer and spherical shell geometries, respectively. The important global diagnostic quantities are the Nusselt number $Nu$ and the Reynolds number $Re$, representing the non-dimensional heat transfer and flow speed.  \\ 

% Heat transfer behaviour: a comparative discussion
An intriguing question in this research direction is the scaling relation between such a diagnostic quantity with a governing input parameter, such as $Nu$, as a function of $Ra$. The thermal convection in planets and stars occurs at parameter values that are several orders of magnitude away from the reach of state-of-the-art numerical simulations and experiments. Therefore, these scaling relations are valuable tools to extrapolate the results of these experiments and simulations to planetary and stellar convective regimes. For plane layer geometry, a Nusselt number scaling of $Nu\thicksim Ra^{2/7}$ is found for moderate thermal forcing ($Ra\leqslant10^{10}$), whereas, for higher thermal forcing, a scaling relation of $Nu\thicksim Ra^{1/3}$ has been widely reported \citep{iyer_2020}. A systematic investigation has been reported by \citep{Gastine2015}, who found the same scaling laws for the Nusselt number in the spherical geometry. It should be noted here that though the global diagnostic quantities exhibit similar behaviour, the local properties, such as the thickness of the viscous and thermal boundary layers, are markedly different in the two geometries. For example, the effect of curvature and a radially varying gravitational acceleration (as appropriate in Earth's core) results in asymmetric boundary layers in the spherical geometry, in contrast to the symmetric boundary layers in a plane layer geometry. \\       

% Basics and motivations for the development of the present solver
Experimental difficulties related to the radial direction of gravity make the advances in spherical shell convection almost entirely dependent on massively parallel numerical simulations. Existing solvers \citep{Wicht2002},\citep{mound_2017} use spherical harmonic decomposition of the flow variables in the azimuthal and latitudinal directions while Chebyshev polynomials are used in the radial direction for proper resolution of the boundary layers. In this paper, we report on the development, implementation, and validation of a new finite-difference solver for studying spherical shell convection. The solver can map any three-dimensional curvilinear geometry to a computational Cartesian domain using the Jacobi transformation. This enables us to solve the conservation equations in Cartesian coordinates, which are much simpler than their spherical coordinate counterpart, even after their modification by the Jacobi, elongation, and stiffness matrix coefficients. Furthermore, the effect of the ellipticity of the core-mantle boundary \citep{forte_1995} and the anisotropic shape of the inner core \citep{yoshida_1996} on the azimuthal and latitudinal variation of radial heat flux can be accounted for. The capability to account for any effect of the non-spherical boundaries is the primary motivation for developing the present solver. The solver uses second-order central spatial discretization, while temporal discretization is achieved with the fractional step method \citep{chongsiripinyo_2019}.  In order to avoid the stiffness induced by the fine resolution near the boundary layers, the viscous terms have been treated implicitly, while the other terms are marched explicitly. The fractional step marches the velocity field into an intermediate field by a combination of the Alternating Direction Implicit method (ADI), the Crank-Nicolson method (CN), and the third-order low-storage Runge-Kutta method (RKW3) \citep{chongsiripinyo_2019}. The remaining procedure in the fractional step method is to remove the divergence residual from the velocity field after the end of each RKW3 step, which in turn is achieved by pressure correction. We use the multigrid HYPRE module to accelerate the pressure correction. The rest of the article is structured as follows. Section \ref{sec:govequ} discusses the governing equation used. The numerical scheme is described in \ref{sec:numerical}. Results are presented in Section \ref{sec:results} and summarized in section \ref{sec:conclusion}.\\

%Basic features of the solver
%The solver has been developed in Computational Fluid Dynamics Lab at the University of California, San Diego, and validated for stratified turbulent flow over a sphere \citep{chongsiripinyo_2019}. Later, the code was customized and validated at the Flow Physics Lab at the Indian Institute of Technology Kanpur for spherical shell convection. 

\section{Governing Equations}\label{sec:govequ}
We aim to investigate Rayleigh-B\'enard convection of an incompressible, Newtonian, Boussinesq fluid in a spherical shell geometry as illustrated in figure \ref{fig:spherical domain}. The spherical shell has an inner radius $r_i$ and an outer radius $r_o$ kept at constant temperatures $T_i$ and $T_o$, respectively. The shell gap $d=r_o-r_i$, the temperature difference $\Delta T=T_i-T_o$, and the free-fall velocity $u_f=\{g_0\alpha(T_i-T_o)d\}^{1/2}$ have been used as the characteristics scale for length, temperature, and velocity, respectively, to nondimensionalize the governing equations. Here, $g_0$ is the gravitational acceleration at the outer radius. The relevant fluid properties are the kinematic viscosity ($\nu$), thermal diffusivity ($\kappa$), and thermal expansion coefficient ($\alpha$). The non-dimensional governing equations are expressed below using a Cartesian coordinate system.\\ 

%\begin{subequation}
\begin{equation}\label{eqn:continuity} 
  \frac{\partial u_j}{\partial x_j} = 0,
\end{equation}
\begin{equation}\label{eqn:momentum} 
  \frac{\partial u_i}{\partial t} + \frac{\partial u_i u_j}{\partial x_j} = -\frac{\partial p}{\partial x_i} + gT \delta_{in} + \sqrt{\frac{Pr}{Ra}} \frac{\partial^2 u_i}{\partial x_j \partial x_j}, 
\end{equation}
\begin{equation}\label{eqn:energy}
  \frac{\partial T}{\partial t} + u_j \frac{\partial T}{\partial x_j}= \frac{1}{\sqrt{RaPr}} \frac{\partial^2 T}{\partial x_j \partial x_j},
\end{equation}
%\end{subequation}
where $g=(r_o/r)^2$ is the radial variation of gravitational acceleration and $\delta_{in}=\cos\theta \;\delta_{i1}+ \sin\theta\cos\phi\;\delta_{i2} + \sin\theta\sin\phi \;\delta_{i3}$. Here $\theta$ and $\phi$ are the colatitude and longitude as shown in figure \ref{fig:spherical domain}. The non-dimensional temperature difference is defined as $T=(T_f-T_o)/(T_i-T_o)$, where $T_f$ is the temperature of the fluid. The non-dimensional parameters in these equations are the Rayleigh number and the Prandtl number defined below.
\begin{equation}\label{eqn:nd_param}
Ra=\frac{g_0 \alpha \Delta T d^3}{\kappa \nu},Pr=\frac{\nu}{\kappa}
\end{equation}
In the subsequent section, we will use a coordinate transformation to convert the spherical domain to a Cartesian domain.

\section{Numerical Algorithms}\label{sec:numerical}
% The numerical algorithms presented in this section have been adapted from \citet{chongsiripinyo_2019}.
\subsection{Coordinate Transformation}\label{sec:coordtrans}

\begin{figure}
\centering
(a)\includegraphics[width=0.34\textwidth, angle=0, clip=true,trim=70mm 8mm 120mm 8mm,scale=1.0]{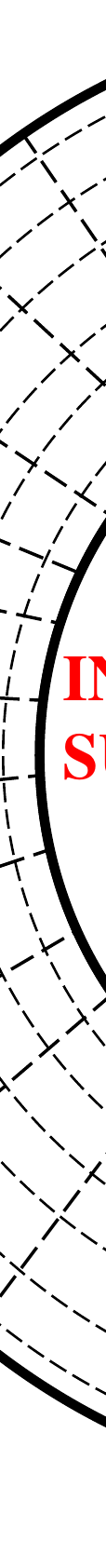}
(b)\includegraphics[width=0.5\textwidth, angle=0, clip=true,trim=60mm 0mm 60mm 0mm,scale=1.0]{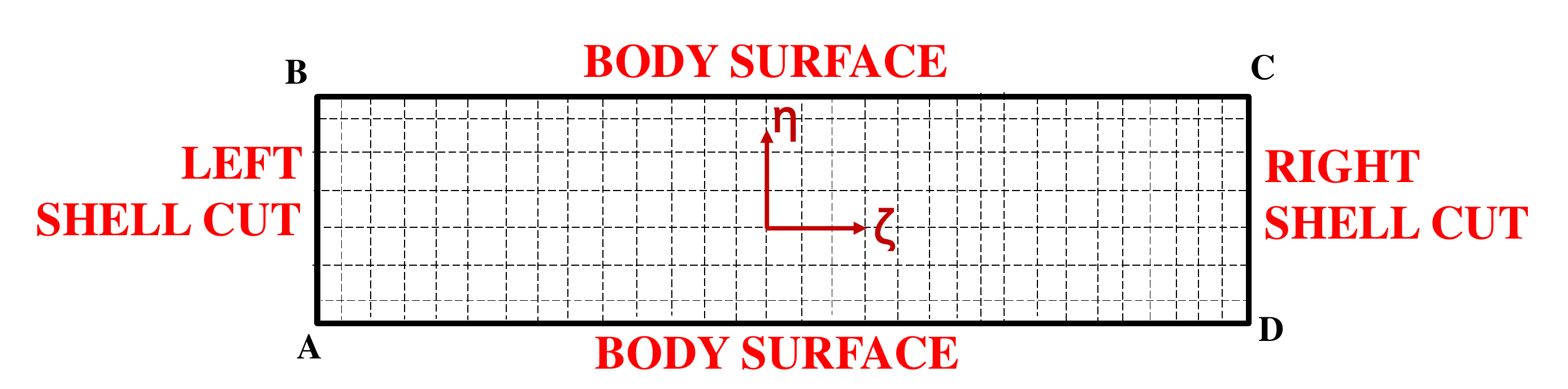}
\caption{(a) Physical curvilinear domain depicting the left half of the $x_3=0$ plane in the spherical geometry with the corresponding (b) transformed Cartesian computational domain.}
\label{fig:domtrans}
\end{figure}

To solve the governing equations \ref{eqn:continuity}-\ref{eqn:energy} in a generalized curvilinear coordinate system we perform coordinate transformation. The basic idea behind a coordinate transformation is to transform a set of physical laws written in Cartesian coordinates $x_1,x_2,x_3$ into an alternative form based on generalized curvilinear coordinates $\zeta, \eta, \xi$ \citep{chongsiripinyo_2019}. \\

\begin{center}
  \fbox{\begin{minipage}{23.4em}
     Physical law written in the Cartesian system \\
      Physical curvilinear grid
  \end{minipage}}
  \begin{center}
     $\Bigg\downarrow \; \text{Grid transformation}$ \\
  \end{center} 
  \fbox{\begin{minipage}{23.4em}
      Physical law written in the generalized system \\
      Computational Cartesian grid, Jacobi terms
  %\caption{Schematic representation of the coordinate transformation or grid transformation strategy used in the present solver}
      
  \end{minipage}}
\end{center}

Such a transformation will result in the inclusion of additional coefficients in the space derivatives in the governing equations, and the relation of this transformation between the Cartesian and the generalized curvilinear coordinate system is stored in a Jacobi matrix ($J$).  The continuity, momentum, and energy equations after the transformation are expressed below.

\begin{equation}\label{eqn:continuity_trans}
\frac{\partial \left[C_{nj} u_j\right]}{\partial \zeta_{n}}=0 
\end{equation}
\begin{equation}\label{eqn:momentum_trans}
\frac{\partial\left|J^{-1}\right| u_i}{\partial t}+\frac{\partial\left[C_{n j} u_j\right] u_i}{\partial \zeta_n}=-\frac{\partial C_{n i} P}{\partial \zeta_n}+gT \delta_{i n}+\sqrt{\frac{Pr}{Ra}}\frac{\partial}{\partial \zeta_n}\left( G_{n j} \frac{\partial u_i}{\partial \zeta_j}\right)
\end{equation}
\begin{equation}\label{eqn:energy_trans}
\frac{\partial T}{\partial t}+\frac{\partial\left[C_{n j}u_j\right]T}{\partial \zeta_n}=\frac{1}{\sqrt{Ra Pr}}\frac{\partial}{\partial \zeta_n}\left(G_{n j}\frac{\partial T}{\partial\zeta_j}\right)    
\end{equation}

Here, $x_i$ denotes the coordinate $i$ of the Cartesian system, and $\zeta_i$ denotes the coordinate $i$ of the generalized system. The notations $x_i=(x_1, x_2, x_3)= (x,y,z)$ and $\zeta_i=(\zeta, \eta, \xi)=(\zeta_1,\zeta_2,\zeta_3)$ have been used interchangeably. After the transformation, the transformed governing equations are solved as if in a Cartesian system. In this context, grid transformation is often synonymously used with coordinate transformation as the curvilinear domain (i.e., a spherical shell domain in our case) is transformed into a new computational Cartesian domain. Here \emph{$J^{-1}$,$C_{i j}$} and \emph{$G_{i j}$} are\\

\begin{equation}\label{eqn:jacobi}
J^{-1}=\left[\begin{array}{lll}
\partial x_1 / \partial \zeta & \partial x_2 / \partial \zeta & \partial x_3 / \partial \zeta\\
\partial x_1 / \partial \eta & \partial x_2 / \partial \eta & \partial x_3 / \partial \eta \\
\partial x_1 / \partial \xi & \partial x_2 / \partial  \xi & \partial x_3 / \partial \xi
\end{array}\right]=:\left[\partial x_i / \partial \zeta_j\right] 
\end{equation}

\begin{equation}\label{eqn:skewness}
C_{i j}=\left|J^{-1}\right| \frac{\partial \zeta_i}{\partial x_j} \quad G_{i j}=\left|J^{-1}\right| \frac{\partial \zeta_i}{\partial x_k} \frac{\partial \zeta_j}{\partial x_k}
\end{equation}

The determinant \emph{$|J^{-1}|$} is the volume ratio of the original cell to the transformed cell, whereas  \emph{$C_{i j}$} and \emph{$G_{i j}$} are grid elongation and skewness coefficients, respectively. The side length, and consequently the side area and the total volume, of a transformed cell is chosen to be unity.\\

% {\color{red}why?}.{\color{blue}This is for convenience. The Jacobian stiffness and elongation matrices are normalized such that in the transformed plane $\Delta\eta=\Delta\zeta=\Delta\xi=1$. Therefore grid spacing does not appear anywhere in the code in the transformed domain. For example, see equation \eqref{PC_term1}. This is common practice.}

%\subsection{Spatial discretization}\label{sec:spatial}

\subsection{Jacobi terms}

\begin{figure}
\begin{minipage}[0.5\linewidth]{0.96\linewidth}
\centering
\includegraphics[width=0.7\textwidth, angle=0, clip=true,trim=0mm 0mm 0mm 0mm,scale=1.0]{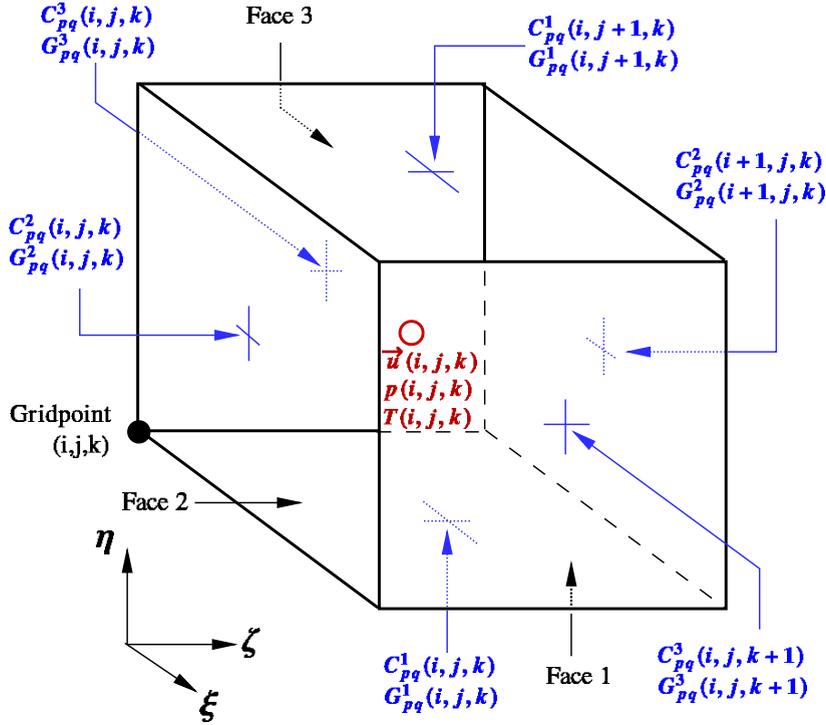}
\end{minipage}
\caption{\label{fig:jacobiBOX} A transformed computational cell associated with the grid point $(i,j,k)$; where $i$, $j$, $k$ are the integer indices used to identify discrete space in the $\zeta$, $\eta$, and $\xi$ directions, respectively.}
\end{figure}

%{\color{red} Shouldn't the Gridpoint (i,j,k) be at the center of the cell?} {\color{blue}In the fibre code, a cell is defined as a box with 8 grid points at the vertices. In Kyle's code, we are using a staggered grid arrangement. Therefore in that code, the cell is defined as a box with staggered gridpoints $(x_e,y_e,z_e)$ at the vertices while the main gridpoints $(x_c,y_c,z_c)$ stay at the centre of the cell. }

Figure \ref{fig:jacobiBOX} demonstrates a cell $(i,j,k)$ in a transformed computational domain. The Jacobi terms, $J^{-1}$, $C_{pq}$, and $G_{pq}$, as expressed in equations \ref{eqn:jacobi} and \ref{eqn:skewness} are stored at the cell's faces. The calculation of  $J^{-1}$ , $C_{pq}$ and $G_{pq}$ is given below.\\

\begin{enumerate}
  \setcounter{enumi}{0}
\item $J^{-1}$ is computed at every cell face, denoted by $J^{-1,fc}$; where $fc$ indicates cell face (1-3), by calculating all the nine components in $J^{-1}$. For instance, we can compute the components of $J^{-1,2}$ of a cell (i,j,k) as follows,\\

  \begin{align*}
    (\partial \overrightarrow{x}/\partial \zeta)|_{i,j,k} = 0.125*(&+ \overrightarrow{x}|_{i+1,j+1,k} + \overrightarrow{x}|_{i+1,j,k} + \overrightarrow{x}|_{i+1,j+1,k+1} + \overrightarrow{x}|_{i+1,j,k+1} \\ &- \overrightarrow{x}|_{i-1,j+1,k} - \overrightarrow{x}|_{i-1,j,k} - \overrightarrow{x}|_{i-1,j+1,k+1} - \overrightarrow{x}|_{i-1,j,k+1})  \\
    (\partial \overrightarrow{x}/\partial \eta)|_{i,j,k} = 0.5*(&+ \overrightarrow{x}|_{i,j+1,k} + \overrightarrow{x}|_{i,j+1,k+1} - \overrightarrow{x}|_{i,j,k} - \overrightarrow{x}|_{i,j,k+1}) \\
    (\partial \overrightarrow{x}/\partial \xi)|_{i,j,k} = 0.5*(&+ \overrightarrow{x}|_{i,j+1,k+1} + \overrightarrow{x}|_{i,j,k+1} - \overrightarrow{x}|_{i,j,k} - \overrightarrow{x}|_{i,j+1,k}).
  \end{align*}
\item Calculate $det(J^{-1,fc})$, denoted by $|J^{-1,fc}|$.
\item The variable $|J^{-1}|$ in equation \ref{eqn:momentum_trans} is an averaged value at the cell center calculated from the six surrounding faces:
  \begin{equation*}
    |J^{-1}|_{i,j,k} = \frac{1}{6}\left(\sum_{fc=1}^{3} |J^{-1,fc}|_{i,j,k} + |J^{-1,1}|_{i,j+1,k} + |J^{-1,2}|_{i+1,j,k}  + |J^{-1,3}|_{i,j,k+1}\right).
  \end{equation*}
\item Compute $J^{fc}=\left[ \partial \zeta_i/\partial x_j  \right]$ simply by the straight-forward inversion, $J^{-1,fc}$: \\ $\{J^{-1,fc}\}^{-1}=det(J^{-1,fc})^{-1}\{cof(J^{-1})\}^T$
\item Calculate $C_{pq}$ and $G_{pq}$ at face $fc$, denoted by $C^{fc}_{pq}$ and $G^{fc}_{pq}$ from $J^{fc}$ using equation \ref{eqn:skewness}.
\end{enumerate}

\subsection{Spatial discretization}\label{sec:spatial}
%\vspace{0.25cm}
\begin{figure}
\begin{minipage}[0.5\linewidth]{0.96\linewidth}
\centering
\includegraphics[width=1.0\textwidth, angle=0, clip=true,trim=0mm 0mm 0mm 0mm,scale=1.025]{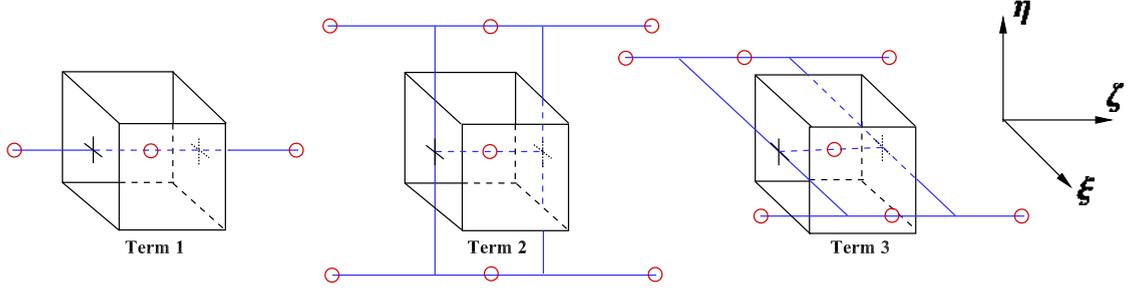}
\end{minipage}
\caption{\label{fig:HypreStencil_Term123} Stencils used for computing \ref{PC_term1}, \ref{PC_term2}, and \ref{PC_term3}.}
\end{figure}

The spatial derivatives in \ref{eqn:momentum_trans} and \ref{eqn:energy_trans}  are discretized using a second-order central finite difference scheme. Figure~\ref{fig:HypreStencil_Term123} illustrates the stencils used to discretize the term \ref{PC_8} using this scheme.\\ 

\begin{equation}
  \frac{\partial}{\partial \zeta_p}  \left[ G_{pq} \frac{\partial \phi}{\partial \zeta_q} \right]
\label{PC_8}
\end{equation}
Equation \ref{PC_8} consists of $9$ terms. We present the discretization of term 1 ($p$=1 and $q$=1), term 2 ($p$=1 and $q$=2), and term 3 ($p$=1 and $q=3$) as examples. %Figure~\ref{fig:HypreStencil_Term123} show stencils used for the examples.
\begin{align}
  \left( \frac{\delta}{\delta \zeta_1}  \left[ G_{11} \frac{\delta \phi}{\delta \zeta_1} \right] \right)_{i,j,k} = &\left[ G_{11} \frac{\delta \phi}{\delta \zeta_1} \right]_{i+1/2,j,k} - \left[ G_{11} \frac{\delta \phi}{\delta \zeta_1} \right]_{i-1/2,j,k} \nonumber \\ = &+G^{2}_{11}|_{i+1,j,k}[\phi|_{i+1,j,k} - \phi|_{i,j,k}] \nonumber \\ &- G^{2}_{11}|_{i,j,k} \quad [\phi|_{i,j,k} - \phi|_{i-1,j,k}]
  \label{PC_term1}  
\end{align}
\begin{align}
  \left( \frac{\delta}{\delta \zeta_1}  \left[ G_{12} \frac{\delta \phi}{\delta \zeta_2} \right] \right)_{i,j,k} = &\left[ G_{12} \frac{\delta \phi}{\delta \zeta_2} \right]_{i+1/2,j,k} - \left[ G_{12} \frac{\delta \phi}{\delta \zeta_2} \right]_{i-1/2,j,k} \nonumber \\ = &+\frac{G^{2}_{12}|_{i+1,j,k}}{2} \left[ + \frac{\phi|_{i,j+1,k} + \phi|_{i+1,j+1,k}}{2} \right. \nonumber \\ &\qquad \qquad \qquad \left. - \frac{\phi|_{i,j-1,k} + \phi|_{i+1,j-1,k}}{2}  \right] \nonumber \\ &-\frac{G^{2}_{12}|_{i,j,k}}{2} \quad \left[ + \frac{\phi|_{i,j+1,k} + \phi|_{i-1,j+1,k}}{2} \right. \nonumber \\ & \quad   \qquad \qquad \: \: \: \: \, \left. - \frac{\phi|_{i,j-1,k} + \phi|_{i-1,j-1,k}}{2}  \right]
\label{PC_term2}
\end{align}
\begin{align}
  \left( \frac{\delta}{\delta \zeta_1}  \left[ G_{13} \frac{\delta \phi}{\delta \zeta_3} \right] \right)_{i,j,k} = &\left[ G_{13} \frac{\delta \phi}{\delta \zeta_3} \right]_{i+1/2,j,k} - \left[ G_{13} \frac{\delta \phi}{\delta \zeta_3} \right]_{i-1/2,j,k} \nonumber \\ = &+\frac{G^{2}_{13}|_{i+1,j,k}}{2}\left[ + \frac{\phi|_{i,j,k+1} + \phi|_{i+1,j,k+1}}{2} \right. \nonumber \\  &\qquad \qquad \qquad \left. - \frac{\phi|_{i,j,k-1} + \phi|_{i+1,j,k-1}}{2}  \right] \nonumber \\ &-\frac{G^{2}_{13}|_{i,j,k}}{2} \quad \left[ + \frac{\phi|_{i,j,k+1} + \phi|_{i-1,j,k+1}}{2} \right. \nonumber \\ & \quad   \qquad \qquad \: \: \: \: \, \left. - \frac{\phi|_{i,j,k-1} + \phi|_{i-1,j,k-1}}{2}  \right] 
  \label{PC_term3}
\end{align}

\subsection{Temporal discretization}\label{sec:temporal}
 For temporal discretization, a fractional step method is used where a velocity field is sequentially advanced in multiple substeps. We use a combination of the Alternating Direction Implicit method (ADI), the Crank-Nicolson method (CN), and the third-order low-storage Runge-Kutta method (RKW3) to march to an intermediate field as described below \citep{chongsiripinyo_2019}.

\subsubsection{Alternating Direction Implicit method}
Alternating Direction Implicit (ADI) method has been used to treat the viscous term implicitly while marching in one direction at a time. We demonstrate the method with a two-dimensional diffusion equation as shown in equation \ref{eqn:adi_1}. To solve \ref{eqn:adi_1} using the Euler method, the procedure is to perform implicit Euler in the $x$ direction with explicit Euler in the $y$ direction for the first half ($\Delta t/2$), and vice versa for the second half $\Delta t/2$ as shown in \ref{eqn:adi_2} and \ref{eqn:adi_3}.

\begin{equation}
  \frac{\partial \phi}{\partial t} = \alpha \left[ \frac{\partial^2 \phi}{\partial x^2} + \frac{\partial^2 \phi}{\partial y^2} \right]
\label{eqn:adi_1}
\end{equation}

\begin{equation}
  \frac{\phi^{n+\frac{1}{2}} - \phi^{n}}{\Delta t/2} = \alpha \left[ \frac{\partial^2 \phi^{n+\frac{1}{2}}}{\partial x^2} + \frac{\partial^2 \phi^n}{\partial y^2} \right]
\label{eqn:adi_2}
\end{equation}

\begin{equation}
  \frac{\phi^{n+1} - \phi^{n+\frac{1}{2}}}{\Delta t/2} = \alpha \left[ \frac{\partial^2 \phi^{n+\frac{1}{2}}}{\partial x^2} + \frac{\partial^2 \phi^{n+1}}{\partial y^2} \right]
\label{eqn:adi_3}
\end{equation}
%\noindent\makebox[\linewidth]{\rule{\textwidth}{0.4pt}}

\subsubsection{Crank-Nicolson method}
The Crank-Nicolson (CN) method splits the right-hand side into two equal parts, the implicit and the explicit, as demonstrated in \ref{cn_1} and \ref{cn_2}. 
\begin{equation}
  \frac{\partial \phi}{\partial t} = \alpha \frac{\partial^2 \phi}{\partial x^2}
\label{cn_1}
\end{equation}

\begin{equation}
  \frac{\phi^{n+1} - \phi^{n}}{\Delta t} = \frac{\alpha}{2} \left[ \frac{\partial^2 \phi^{n+1}}{\partial x^2} + \frac{\partial^2 \phi^n}{\partial x^2} \right]
\label{cn_2}
\end{equation}
%\noindent\makebox[\linewidth]{\rule{\textwidth}{0.4pt}}

\subsubsection{Third order Runge-Kutta method}

\begin{table}
\begin{center}
\vspace{-0.5 cm}
\begin{tabular}{c c c c}
& & &\\ % put some space after the caption
\hline
$\quad$ Substep $\quad$ &$\quad$ $h$ $\quad$ & $\quad$ $\beta$ $\quad$ &$\quad$  $\Pi$ $\quad$  \\ \hline
   1    & 8$\Delta t$/15  &    1        &     0         \\
   2    & 2$\Delta t$/15  &  25/8       & -17/8 \\
   3    & 1$\Delta t$/3   &   9/4       & -5/4 \\
\hline 
\end{tabular}
\caption{RKW3 parameters.}
\label{tab:rkw3}
\end{center}
\end{table}
%%%%%%%%%%%%%%%% end table %%%%%%%%%%%%%%%%%%%

The third-order low-storage Runge-Kutta method (RKW3) uses only two storage variables. Marching is accomplished in three substeps, briefly summarized here. Given an equation for $\phi$, 
\begin{equation}
  \frac{\partial \phi}{\partial t} = \text{R}(\phi).
\label{RKW3_1}
\end{equation}
RKW3 is implemented in the following manner, 
\begin{equation}
  \frac{\phi^{rk} - \phi^{rk-1}}{{h^{rk}}} = {\beta^{rk}}\text{R}(\phi^{rk-1}) + {\Pi^{rk}}\text{R}(\phi^{rk-2}) \, .
\label{RKW3_2}
\end{equation}
Here, $rk$ goes from substep 1 to substep 3, and the values of ${h}$, ${\beta}$, and ${\Pi}$ are given in table \ref{tab:rkw3}.

\subsubsection{The ADI-CN-RKW3 combined marching scheme}\label{sec:timemarching}

\begin{figure}
\centering
\includegraphics[scale=.65]{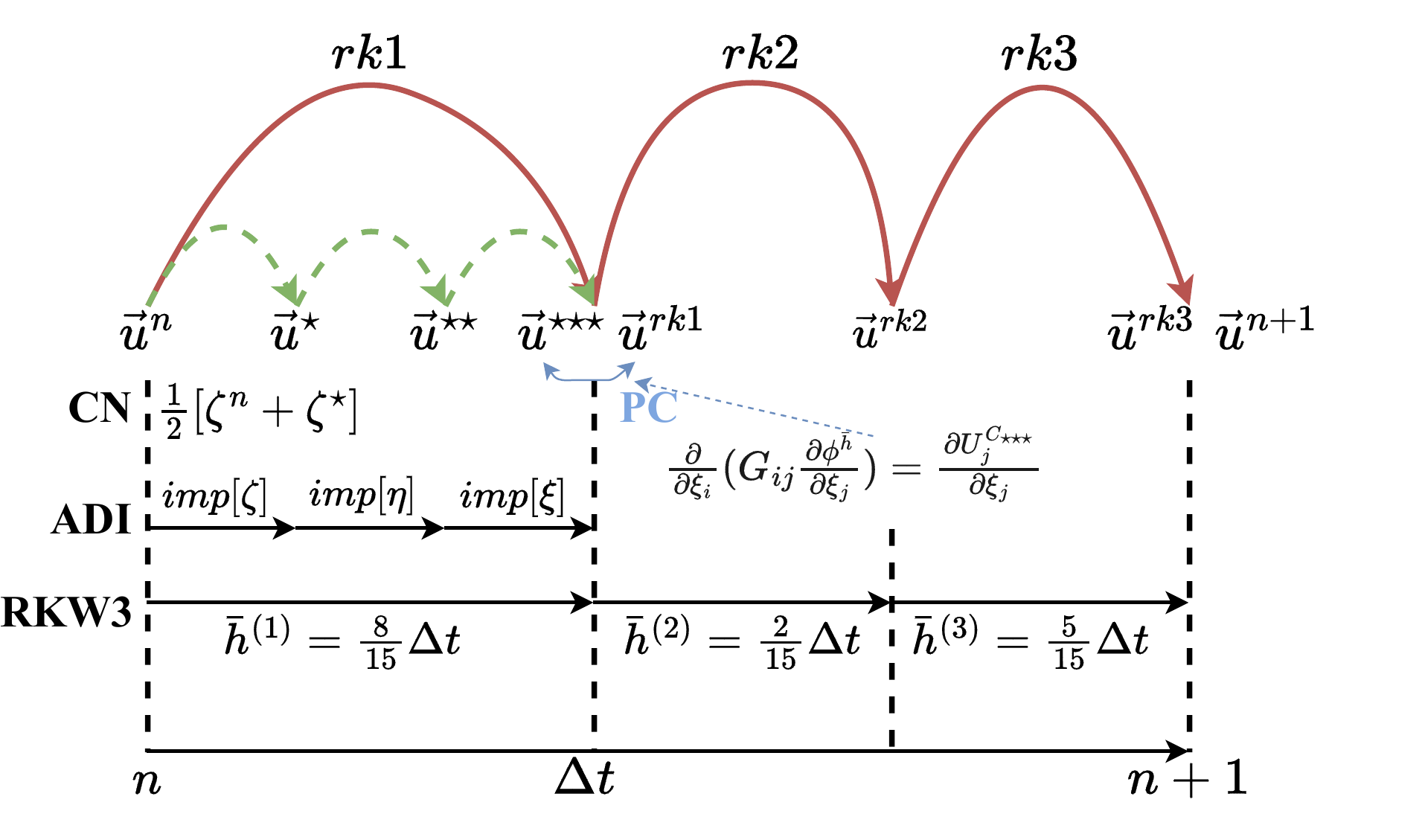}
\caption{ADI-CN-RKW3 combined marching scheme. PC denotes the pressure correction.}
\label{fig:ADI CN RKW3}
\end{figure}

The above-mentioned algorithms (ADI, CN, and RKW3) are combined to march the governing equations to an intermediate state temporally. The right-hand side of equation \ref{eqn:momentum_trans} is split into explicit and implicit terms as indicated by the subscripts $ex$ and $im$ in equation \ref{solve_1}. Depending on the grid skewness $G_{ij}$, the diagonal components of the viscous terms are susceptible to the stiffness of the discretized systems and are, therefore, marched implicitly. The ADI scheme is used since there are three viscous terms containing $G_{11}$, $G_{22}$, and $G_{33}$. At a given time, they are split into two parts using CN. These steps are shown in \ref{solve_2}, \ref{solve_3}, and \ref{solve_4} as an example for substep 1 of the RKW3 marching scheme. \\

%{\color{red}Why are we using such a combination, would it stabilize the code in terms of the time step size?}{\color{blue}Yes. We are using the implicit scheme for the viscous term to increase the time step size. ADI is convenient as we do not need to consider all the viscous terms implicitly at once. Fully implicit and Crank-Nicolson are both unconditionally stable schemes, but CN is more accurate. Overall in this fractional step method, we increase the time step size at the expense of third-order accuracy of purely explicit RKW3}\\

\begin{eqnarray}
\begin{split}
\frac{\partial J^{-1} u_i}{\partial t}=\left[-\frac{\partial C_{n i} P}{\partial \zeta_n}-\frac{\partial\left[C_{n j} u_j\right] u_i}{\partial \zeta_n}+\sqrt{\frac{Pr}{Ra}}\left(\frac{\partial}{\partial \zeta_n} G_{n j} \frac{\partial u_i}{\partial \zeta_j}\right)_{n\neq j}+gT \delta_{i n}\right]_{ex} \\
+\left[\sqrt{\frac{Pr}{Ra}}\left(\frac{\partial}{\partial \zeta_n} G_{n j} \frac{\partial u_i}{\partial \zeta_j}\right)_{n=j}\right]_{i m}
\label{solve_1}
\end{split}
\end{eqnarray}

\begin{eqnarray}
%\centering
\begin{split}
J^{-1} u_i^{\star}=J^{-1} u_i^n+\beta^{(1)} h^{(1)} \square +\sqrt{\frac{Pr}{Ra}}\frac{h^{(1)}}{2}\left(\frac{\partial}{\partial \zeta}\left[G_{11} \frac{\partial u_i^n}{\partial \zeta}\right]+\frac{\partial}{\partial \zeta}\left[G_{11} \frac{\partial u_i^{\star}}{\partial \zeta}\right]\right),\\
+\sqrt{\frac{Pr}{Ra}} h^{(1)} \frac{\partial}{\partial \eta}\left[G_{22} \frac{\partial u_i^n}{\partial \eta}\right]+\sqrt{\frac{Pr}{Ra}} h^{(1)} \frac{\partial}{\partial \xi}\left[G_{33} \frac{\partial u_i^n}{\partial \xi}\right],
\label{solve_2}
\end{split}   
\end{eqnarray}
Here, $\square$ represents all the terms to be marched explicitly.

\begin{eqnarray}
\begin{split}
J^{-1} u_i^{\star \star}=J^{-1} u_i^{\star}-\sqrt{\frac{Pr}{Ra}} \frac{h^{(1)}}{2} \frac{\partial}{\partial \eta}\left[G_{22} \frac{\partial u_i^n}{\partial \eta}\right]+\sqrt{\frac{Pr}{Ra}} \frac{h^{(1)}}{2} \frac{\partial}{\partial \eta}\left[G_{22} \frac{\partial u_i^{\star \star}}{\partial \eta}\right] 
\label{solve_3}
\end{split}
\end{eqnarray}

\begin{eqnarray}
\begin{split}
J^{-1} u_i^{\star \star \star}=J^{-1} u_i^{\star \star}-\sqrt{\frac{Pr}{Ra}} \frac{h^{(1)}}{2} \frac{\partial}{\partial \xi}\left[G_{33} \frac{\partial u_i^n}{\partial \xi}\right]+\sqrt{\frac{Pr}{Ra}}\frac{h^{(1)}}{2} \frac{\partial}{\partial \xi}\left[G_{33} \frac{\partial u_i^{\star \star\star}}{\partial \xi}\right] 
\label{solve_4}
\end{split}
\end{eqnarray}

The intermediate velocity fields $u^{\star}_i$, $u^{\star \star}_i$, $u^{\star \star \star}_i$ are obtained by solving a set of tridiagonal matrices that result from the spatial discretization of the equations \ref{solve_2}, \ref{solve_3}, and \ref{solve_4} in the $\zeta$, $\eta$, and $\xi$ directions respectively. The intermediate velocity $u^{\star \star \star}_i$ is the first step in the fractional-step scheme. We employ the Thomas Algorithm with Pipelining, as described in the next section, to solve the tridiagonal system \ref{solve_2}-\ref{solve_4} to obtain $u^{\star}$, $u^{\star \star}$, and $u^{\star \star \star}$. \\

\subsection{Thomas Algorithm}\label{sec:thomas}

%A parallel Thomas algorithm has been used along with pipelining (either with or without periodic boundary conditions) to solve the tridiagonal system as described before. It is a remarkably efficient algorithm used to solve a tridiagonal system such that those constructed from \ref{solve_2}, \ref{solve_3}, and \ref{solve_4} for $u^{\star}$, $u^{\star \star}$, and $u^{\star \star \star}$. 

Let us consider solving $A\psi = g$ for $\psi$, which is the outcome of the spatial discretization of, for instance, \ref{solve_2}. Here $A$ is a tridiagonal matrix given as 

%Consider a system $Ax=g$ where A is a tridiagonal matrix:
\begin{equation}
  \left[\begin{array}{ccccccccc}
      b_{0~~} & c_{0~~}                  &&&&&&& \\
      a_{1~~} & b_{1~~} & c_{1~~}        &&&&&& \\
      & a_{2~~} & b_{2~~} & c_{2~~}      &&&&& \\ 
      && \cdot{~~~} & \cdot{~~~} & \cdot{~~~}           &&&& \\ 
      &&& \cdot{~~~} & \cdot{~~~} & \cdot{~~~}          &&& \\ 
      &&&& a_{n-1}    & b_{n-1} & c_{n-1} &&  \\
      &&&&& a_{n~~}    & b_{n~~}  \\
    \end{array}\right]  
  \left[{\begin{array}{c}
        \psi^{~}_{0~~} \\
        \psi^{~}_{1~~} \\
        \psi^{~}_{2~~} \\
        ~\cdot_{~~~}  \\
        ~\cdot_{~~~}  \\
        \psi^{~}_{n-1} \\
        \psi^{~}_{n~~} \\
    \end{array}}\right]
  =\left[{\begin{array}{c}
        g^{~}_{0~~} \\
        g^{~}_{1~~} \\
        g^{~}_{2~~} \\
        ~\cdot_{~~~}  \\
        ~\cdot_{~~~}  \\
        g^{~}_{n-1} \\
        g^{~}_{n~~} \\
    \end{array}}\right].
    \label{eqn:FIBRE_TA1}
\end{equation}

\begin{flushleft}
  The first two relations in \ref{eqn:FIBRE_TA1} are,
\end{flushleft}
\begin{align}
  b_0 \psi_0 + c_0 \psi_1 &= g_0 \label{TA1_line1} \\
  a_1 \psi_0 + b_1 \psi_1 + c_1 \psi_2 &= g_1. \label{TA1_line2}
\end{align}
Substituting $\psi_0$ from \ref{TA1_line1} into $\psi_0$ in \ref{TA1_line2} gives
%% \begin{equation}
%%   a_1 \left[ \dfrac{g_0 - c_0 x_1}{b_0} \right] + b_1 x_1 + c_1 x_2 = g_1
%% \end{equation}
\begin{equation}    
  b^{'}_1 \psi_1 + c_1 \psi_2 = g^{'}_1
\end{equation}
where $b^{'}_1 = \left[ b_1 - a_1 b^{-1}_0 c_0  \right]$ and $g^{'}_1 = \left[ g_1 - a_1 b^{-1}_0 g_0  \right]$. The algorithm involves two stages, forward sweeping and backward substitution. The sub-diagonal elements $a_1-a_n$ are removed using Gaussian elimination during the forward sweeping step. Therefore, equation \ref{eqn:FIBRE_TA1} takes the form:
\begin{equation}
  \left[\begin{array}{ccccccccc}
      b_{0~~} & c_{0~~}                  &&&&&&& \\
      0    & b^{'}_{1~~} & c_{1~~}        &&&&&& \\
      &  0    & b^{'}_{2~~} & c_{2~~}      &&&&& \\ 
      && \cdot{~~~} & \cdot{~~~} & \cdot{~~~}           &&&& \\ 
      &&& \cdot{~~~} & \cdot{~~~} & \cdot{~~~}          &&& \\ 
      &&&&      0     & b^{'}_{n-1} & c_{n-1} &&  \\
      &&&&&      0              & b^{'}_{n~~} \\
    \end{array}\right]  
  \left[{\begin{array}{c}
        \psi^{~}_{0~~} \\
        \psi^{~}_{1~~} \\
        \psi^{~}_{2~~} \\
        ~\cdot_{~~~}  \\
        ~\cdot_{~~~}  \\
        \psi^{~}_{n-1} \\
        \psi^{~}_{n~~} \\
    \end{array}}\right]
  =\left[{\begin{array}{c}
        g^{~}_{0~~} \\
        g^{'}_{1~~} \\
        g^{'}_{2~~} \\
        ~\cdot_{~~~}  \\
        ~\cdot_{~~~}  \\
        g^{'}_{n-1} \\
        g^{'}_{n~~} \\
    \end{array}}\right].
\label{eqn:FIBRE_TA2}
\end{equation}  
At the end of the forward sweep, we can solve for $\psi_n = g^{'}_n/b^{'}_n$ in \ref{eqn:FIBRE_TA2}.  Subsequently, we solve for $\psi_{n-1}$-$\psi_0$ (equations $n-1$ until $0$) as in $\psi_i = (g^{'}_i - c_i \psi_{i+1})/b^{'}_i$. For the grid in Section~\ref{sec:details}, periodic boundary conditions are enforced in the $\xi$ direction. Therefore, the above-mentioned Thomas algorithm is modified as follows. Consider the discretized system $A\psi = g$ with periodicity as in \ref{eqn:FIBRE_TAP1} where $\psi_1=\psi_{n-1}$ and $\psi_2=\psi_{n}$.

\begin{equation}
  \left[\begin{array}{cccccccc}
      b_{1~~} & c_{1~~}        &&&&& a_{1~~} \\
      a_{2~~} & b_{2~~} & c_{2~~}      &&&& \\ 
      & \cdot{~~~} & \cdot{~~~} & \cdot{~~~}   &&& \\ 
      && \cdot{~~~} & \cdot{~~~} & \cdot{~~~}     && \\ 
      &&& \cdot{~~~} & \cdot{~~~} & \cdot{~~~}     &     \\
      &&&&  a_{n-1} & b_{n-1} & c_{n-1}   \\
      c_{n~~} &&&&& a_{n~~}    & b_{n~~} \\
    \end{array}\right]  
  \left[{\begin{array}{c}
        \psi^{~}_{1~~} \\
        \psi^{~}_{2~~} \\
        ~\cdot_{~~~}  \\
        ~\cdot_{~~~}  \\        
        ~\cdot_{~~~}  \\
        \psi^{~}_{n-1} \\        
        \psi^{~}_{n~~} \\
    \end{array}}\right]
  =\left[{\begin{array}{c}
        g^{~}_{1~~} \\
        g^{~}_{2~~} \\
        ~\cdot_{~~~}  \\
        ~\cdot_{~~~}  \\        
        ~\cdot_{~~~}  \\
        g^{~}_{n-1} \\        
        g^{~}_{n~~} \\
    \end{array}}\right]
  \label{eqn:FIBRE_TAP1}
\end{equation} 
The first step includes separating equation \ref{eqn:FIBRE_TAP1} into a tridiagonal system \ref{eqn:FIBRE_TAP2} with an additional equation \ref{eqn:FIBRE_TAP3}.
\begin{equation}
  \left[\begin{array}{ccccccc}
      b_{1~~} & c_{1~~}        &&&& \\
      a_{2~~} & b_{2~~} & c_{2~~}      &&& \\ 
      & \cdot{~~~} & \cdot{~~~} & \cdot{~~~}   && \\ 
      && \cdot{~~~} & \cdot{~~~} & \cdot{~~~}     & \\ 
      &&& \cdot{~~~} & \cdot{~~~} & \cdot{~~~}          \\
      &&&&  a_{n-1} & b_{n-1}   \\
    \end{array}\right]  
  \left[{\begin{array}{c}
        \psi^{~}_{1~~} \\
        \psi^{~}_{2~~} \\
        ~\cdot_{~~~}  \\        
        ~\cdot_{~~~}  \\
        ~\cdot_{~~~}  \\        
        \psi^{~}_{n-1} \\
    \end{array}}\right]
  =\left[{\begin{array}{c}
        g^{~}_{1~~} \\
        g^{~}_{2~~} \\
        ~\cdot_{~~~}  \\        
        ~\cdot_{~~~}  \\
        ~\cdot_{~~~}  \\        
        g^{~}_{n-1} \\
    \end{array}}\right]
  +\left[{\begin{array}{c}
        -a^{~}_{1~~} \\
        ~0^{~}_{~~~} \\
        ~0^{~}_{~~~} \\
        ~\cdot_{~~~}  \\
        ~\cdot_{~~~}  \\        
        -c^{~}_{n-1} \\
    \end{array}}\right] \psi_n
  \label{eqn:FIBRE_TAP2}
\end{equation} 

\begin{equation}
  c_n \psi_1 + a_n \psi_{n-1} + b_n \psi_n = g_n.
  \label{eqn:FIBRE_TAP3}
\end{equation} 
The tridiagonal matrix on the left-hand side is defined as $[A_c]$ and $[g]$ as the g-column matrix on the right-hand side. Let
\begin{equation}
  [\psi] = [\psi1] + [\psi2]\psi_n
  \label{eqn:FIBRE_TAP4}
\end{equation} 
be the solution of the system \ref{eqn:FIBRE_TAP2} where
\begin{align}
  [\psi1] &= [A_c]^{-1}[g] \label{eqn:FIBRE_TAP5} \\
  [\psi2] &= [A_c]^{-1}[-a_1 0 \cdot \cdot \cdot -c_{n-1}]^T.   \label{eqn:FIBRE_TAP6}
\end{align} 
Substituting $\psi$ in \ref{eqn:FIBRE_TAP4} into $\psi_1$ and $\psi_{n-1}$ in \ref{eqn:FIBRE_TAP3} gives
\begin{equation}
  c_n(\psi1_1 + \psi2_1 \psi_n) + a_n(\psi1_{n-1} + \psi2_{n-1} \psi_n) + b_n \psi_n = g_n.
  \label{eqn:FIBRE_TAP7}
\end{equation} 
Rearrange \ref{eqn:FIBRE_TAP7} for $\psi_n$
\begin{equation}
  \psi_n = \dfrac{g_n - c_n \psi1_1 - a_n \psi1_{n-1}}{b_n + c_n \psi2_1 + a_n \psi2_{n-1}}.
  \label{eqn:FIBRE_TAP8}
\end{equation}

In summary, to solve the system \ref{eqn:FIBRE_TAP1}, we employ the following steps: 
\begin{enumerate}
\item Construct \ref{eqn:FIBRE_TAP5} and \ref{eqn:FIBRE_TAP6}
\item Solve \ref{eqn:FIBRE_TAP5} and \ref{eqn:FIBRE_TAP6} for $[\psi1]$ and $[\psi2]$ from index 1 to index $n-1$
\item Substitute $[\psi1]$ and $[\psi2]$ into \ref{eqn:FIBRE_TAP8} and solve for $\psi_n$
\item Calculate $[\psi]$ from \ref{eqn:FIBRE_TAP4} using $[\psi1]$, $[\psi2]$, and $\psi_n$
\end{enumerate}  

\subsubsection{Parallel algorithm}

%\newpage
\begin{equation}
  \begin{array}{c}
    \phantom{a} \\ 
    \phantom{a} \\ 
    \phantom{a} \\
            {\color{blue} \rotatebox{90}{$\text{CPU}^{i}$} } \\
            \phantom{a} \\ 
            \phantom{a} \\
            {\color{red} \rotatebox{90}{$\text{CPU}^{i+1}$} } \\
            \phantom{a} \\ 
            \phantom{a} \\
  \end{array} 
  \left| \begin{array}{ccccccccccc | c} 
      {\color{olive} b^{'i-1}_{s~~}} & {\color{olive} c^{i-1}_{s~~}}  &&&&&&&&&& \cdot{~~~} \\ \hline
      0 &b^{'i~~}_{s~~} & c^{i~~}_{s~~} &&&&&&&&& g^{'i~~}_{s~~} \\ 
      & 0 & b^{'i~~}_{s+1} & c^{i~~}_{s+1} &&&&&&&& g^{'i~~}_{s+1} \\ 
      && \cdot{~~~} & \cdot{~~~} & \cdot{~~~} &&&&&&& \cdot{~~~} \\ 
      &&& 0 & {\color{blue} b^{'i~~}_{e~~}} & {\color{blue} c^{i~~}_{e~~}} &&&&&& {\color{blue} g^{'i~~}_{e~~}} \\ \hline
      &&&&  a^{i+1}_{s~~}   & b^{i+1}_{s~~} & c^{i+1}_{s~~} &&&&& {\color{red} g^{i+1}_{s~~}} \\ 
      &&&&& a^{i+1}_{s+1} & b^{i+1}_{s+1} & c^{i+1}_{s+1} &&&&  g^{i+1}_{s+1}  \\ 
      &&&&&& \cdot{~~~} & \cdot{~~~} &  \cdot{~~~} &&& \cdot{~~~}  \\
      &&&&&&& a^{i+1}_{e~~} & {\color{red} b^{i+1}_{e~~}} & {\color{red} c^{i+1}_{e~~}} && {\color{red} g^{i+1}_{e~~}} \\ \hline
      &&&&&&&& {\color{red} a^{i+1}_{s~~}} & \cdot{~~~} & \cdot{~~~} & \cdot{~~~} \\       
  \end{array} \right|
  \label{eqn:FIBRE_PTAF}  
\end{equation} 
\begin{equation}\label{eqn:FIBRE_PTAB}
  \begin{array}{c}
    \phantom{a} \\ 
    \phantom{a} \\ 
    \phantom{a} \\
            {\color{blue} \rotatebox{90}{$\text{CPU}^{i}$} } \\
            \phantom{a} \\ 
            \phantom{a} \\
            {\color{red} \rotatebox{90}{$\text{CPU}^{i+1}$} } \\
            \phantom{a} \\ 
            \phantom{a} \\
  \end{array} 
  \left| \begin{array}{ccccccccccc | c} 
      b^{'i-1}_{s~~} & c^{i-1}_{s~~} &&&&&&&&&& \cdot{~~~} \\ \hline
      &b^{'i~~}_{s~~} & c^{i~~}_{s~~} &&&&&&&&& g^{'i~~}_{s~~} \\ 
      &  & b^{'i~~}_{s+1} & c^{i~~}_{s+1} &&&&&&&& g^{'i~~}_{s+1} \\ 
      &&& \cdot{~~~} & \cdot{~~~} &&&&&&& \cdot{~~~} \\ 
      &&&& b^{'i~~}_{e~~} & c^{i~~}_{e~~} &&&&&& g^{'i~~}_{e~~} \\ \hline
      &&&&  & b^{'i+1}_{s~~} & c^{i+1}_{s~~} &&&&& {\color{red} \psi^{i+1}_{s~~}} \\ 
      &&&&&  & b^{'i+1}_{s+1} & c^{i+1}_{s+1} &&&&  \psi^{i+1}_{s+1}  \\ 
      &&&&&&& \cdot{~~~} &  \cdot{~~~} &&& \cdot{~~~}  \\
      &&&&&&&  & b^{'i+1}_{e~~} & c^{i+1}_{e~~} &&  \psi^{i+1}_{e~~} \\ \hline
      &&&&&&&&& \cdot{~~~} & \cdot{~~~} \\       
  \end{array} \right|
\end{equation} 

\begin{figure}[ht]
\begin{minipage}[0.5\linewidth]{0.96\linewidth}
\centering
\includegraphics[width=1.0\textwidth, angle=0, clip=true,trim=0mm 0mm 0mm 0mm,scale=0.6]{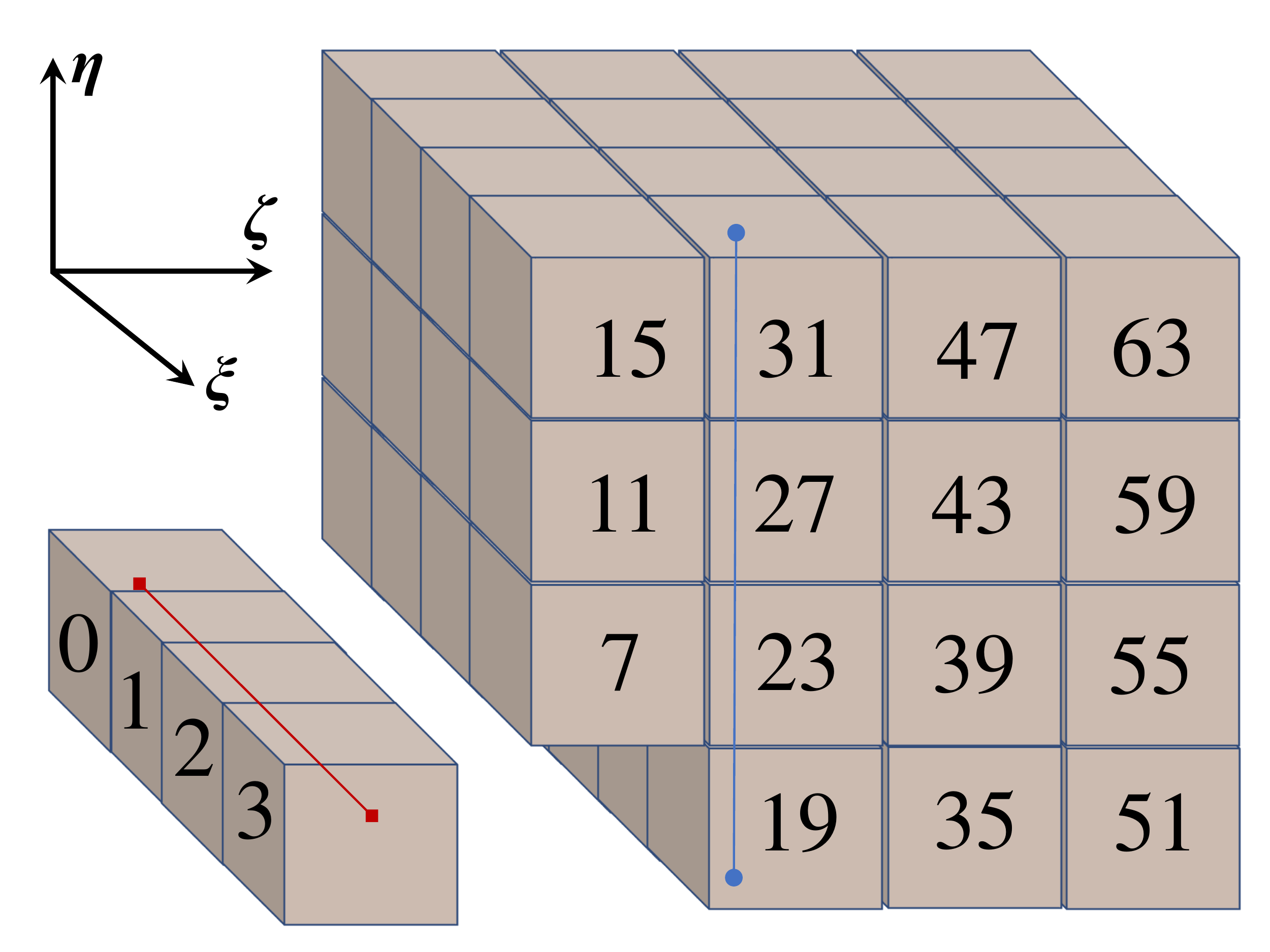}
\end{minipage}
\caption{\label{fig:ThomasParallelled} Example of a 4$\times$4$\times$4 CPU topology in a single computational domain. Numbers in the figure represent an individual CPU's rank (id).}
\end{figure}

Figure \ref{fig:ThomasParallelled} demonstrates an example 4$\times$4$\times$4 central processing unit (CPU) topology in a single computational domain. The spatial discretization of \ref{solve_2}, \ref{solve_3}, or \ref{solve_4} over several CPUs results in tridiagonal matrices, and vectors. The tridiagonal matrices constructed from the discretization of \ref{solve_4} span over the entire $\xi$ space index, e.g. over CPU$^{0-3}$ illustrated by the solid red line in the figure \label{fig:ThomasParalleled}. Similarly, the tridiagonal matrices constructed from the discretization of \ref{solve_3} span over the entire $\eta$ space index (e.g. over CPU$^{19-31}$), as indicated by a solid blue line. Considering \ref{eqn:FIBRE_PTAF}, forward sweeping starts at CPU$^0$. Once the sweeping reaches the interface between CPU$^{i}$ and CPU$^{i+1}$, CPU$^{i}$ sends $b^{'i}_e$, $c^{i}_e$, and $g^{'i}_e$ to CPU$^{i+1}$. Then, CPU$^{i+1}$ continues to carry out the sweeping by sending data $b^{'i+1}_e$, $c^{i+1}_e$, and $g^{'i+1}_e$ to CPU$^{i+2}$ and so on. After the forward sweeping is finalized, backward substitution starts, and reverse sweeping is performed, as shown in \ref{eqn:FIBRE_PTAB}. However, the only information being sent from CPU$^{i+1}$ to CPU$^{i}$ is $\psi^{i+1}_s$. \\

\begin{flushleft}
For a periodic system, we use the following steps:\\
\end{flushleft}
\begin{enumerate}
\item Construct \ref{eqn:FIBRE_TAP5} and \ref{eqn:FIBRE_TAP6}
\item Use the parallel Thomas subroutine to solve \ref{eqn:FIBRE_TAP5} and \ref{eqn:FIBRE_TAP6} for $[\psi1]$ and $[\psi2]$
\item CPU$^0$ owning the first block (contains node 1), sends $\psi1_1$ and $\psi2_1$ to CPU$^N$ that owns the last block (contains node n)
\item CPU$^N$ calculates $\psi_n$ and broadcasts $\psi_n$ to every CPU that owns a subsystem of \ref{eqn:FIBRE_TAP1}
\item Every CPU calculates $[\psi]$ from \ref{eqn:FIBRE_TAP4}
\end{enumerate}  
Notice that by splitting \ref{eqn:FIBRE_TAP1} into \ref{eqn:FIBRE_TAP2} and \ref{eqn:FIBRE_TAP3}, CPU$^N$ solves the tridiagonal system \ref{eqn:FIBRE_TAP4} which has size one element less than the others.

\subsubsection{Pipelining}\label{subsec:Pipelining}

In the previous subsection, we summarize how to solve a tridiagonal system in parallel. It is done simply by completing the forward/backward sweep and sending data to the proper neighbor in order to continue marching. CPU$^i$ that finishes the forward sweep sends data to CPU$^{i+1}$ until the last block is reached. Generally, each CPU can be responsible for thousands of tridiagonal subsystems contained in a single subdomain (or `a block'). This subsection summarizes how to solve such a big system efficiently. \\

Consider a computational domain containing $(n_\zeta,n_\eta,n_\xi)$ grid points. For the sake of simplicity, the domain is equally decomposed only in the $\eta$ direction into $NJ$ blocks so that CPU$^{0}$ occupies block $0$, CPU$^{1}$ occupies block $1$, and so on. Thus, each CPU owns a block of size $(n_\zeta,n_\eta/N_\eta,n_\xi)$; given that $n_\eta/N_\eta$ is, by design, an integer. Supposing that we choose to perform an implicit marching in the $\eta$ direction, the resulting tridiagonal matrix is subdivided into $N_\eta$ sections. The easiest, though the least efficient, way to solve these systems is to let CPU$^0$ solve all of its subsystems across the $(n_\zeta,n_\eta/N_\eta,n_\xi)$ grid before sending data to CPU$^1$. That is, CPU$^0$ performs a forward sweep at cell $(1,1 \rightarrow n_\eta/N_\eta, 1)$, at cell $(1,1 \rightarrow n_\eta/N_\eta, 2)$, and so on until cell $(n_\zeta,1 \rightarrow n_\eta/N_\eta, n_\xi)$. Next, CPU$^0$ packs the plane data with $n_\zeta*n_\xi*3$ elements (recall $b^{'i}_e$, $c^{i}_e$, and $g^{'i}_e$ in the previous subsection) at $(1:n_\zeta, n_\eta/N_\eta, 1:n_\xi)$ and sends it to CPU$^{1}$. Following the same process for the subsequent CPUs until CPU$^{N_{\eta}-1}$ is reached, the backward substitution is carried out in the same way from CPU$^{N_\eta-1}$ to CPU$^{0}$. The obvious drawback is that only one CPU operates at a given time, and the whole process will be even slower than the serial version since there is additional communication overhead.\\ 

\begin{figure}[ht]
\begin{minipage}[0.5\linewidth]{0.96\linewidth}
\centering
\includegraphics[width=0.8\textwidth, angle=0, clip=true,trim=20mm 20mm 20mm 20mm,scale=1.1]{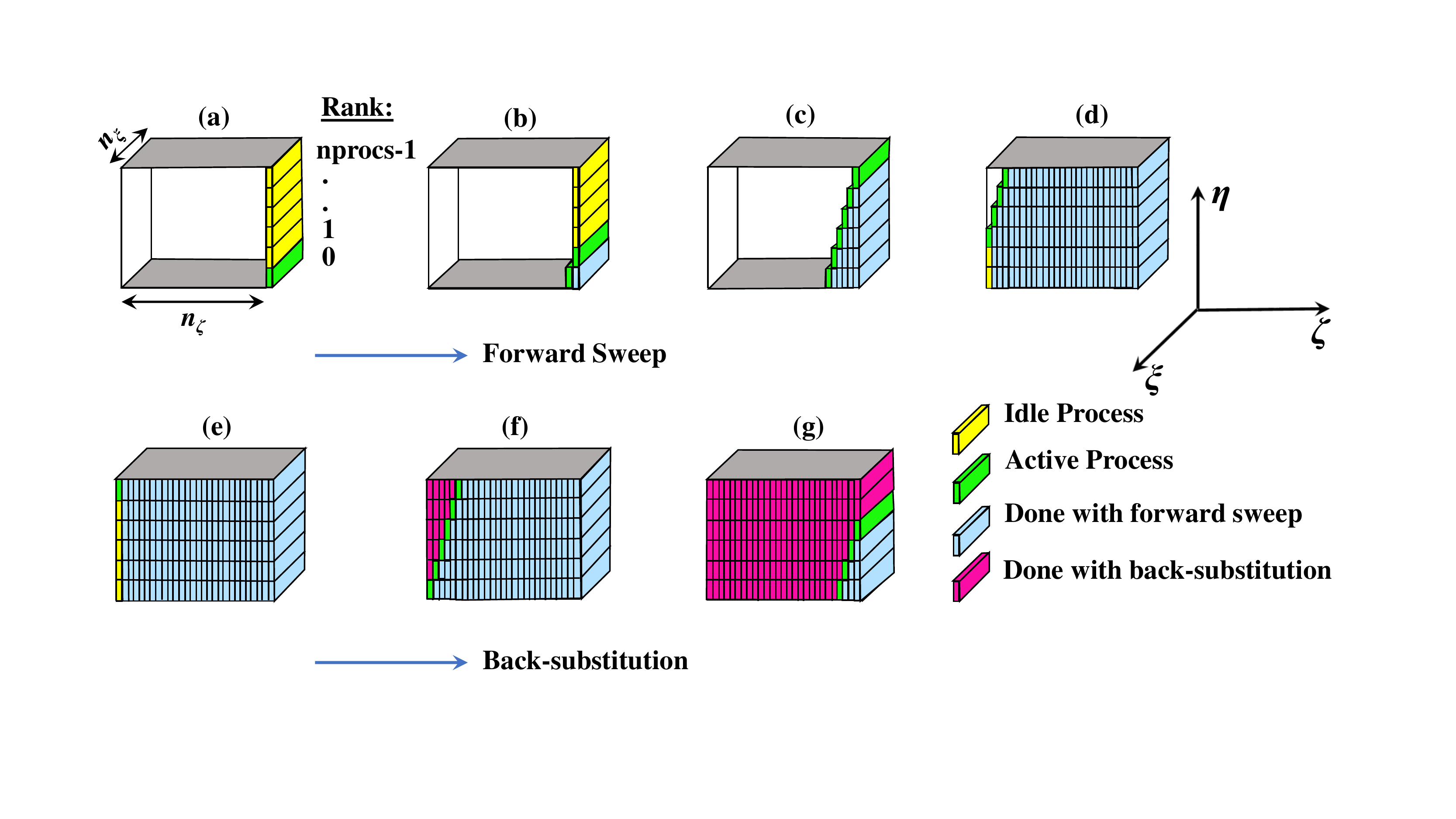}
\end{minipage}
\caption{Illustration of pipelining with the parallel Thomas algorithm following Figure VI.16 of \citet{Taylor2008}. Forward sweep is achieved in steps (a-d), while steps (e-g) depict back-substitution.}
\label{fig:pipelining}
\end{figure}

Pipelining is employed in an attempt to minimize the number of idle CPUs while optimizing communication overhead. In essence, rather than sweeping across the $(n_\zeta,n_\eta/N_\eta,n_\xi)$ grid all at once, each CPU performs the sweeps only for a portion of the grid and shares data with its neighbouring CPU in the sweep direction downstream in a forward sweep and upstream for a backward sweep. A portion of the grid can be chosen for the first CPU, with the others obeying the same portion. We give an example of pencil-type pipelining. Consider figure~\ref{fig:pipelining} and the following steps:
\begin{enumerate}%[label=(\alph*)]
\item CPU$^{0}$, process rank 0 in the figure, performs the forward sweep in the $\eta$ direction at cell $(i,k)=(1,1)$ from $(i,j,k)=(1,1,1)$ to $(i,j,k)=(1,n_\eta/N_\eta,1)$; here $i$ and $k$ are dummy indices pointing to a grid location in $\zeta$ and $\xi$ directions, respectively. CPU$^{0}$ then repeats the forward sweep until $(i,k)=(1,n_\xi)$. Notice that the forward sweep is in the $\eta$ direction, but the `pencil' aligns in the $\xi$ direction. At this point, CPU$^{0}$ packs and passes data to CPU$^{1}$. The data is of size $n_\xi*3$ elements containing $b^{'0}_{n_\eta/N_\eta}$, $c^{0}_{n_\eta/N_\eta}$, and $g^{'0}_{n_\eta/N_\eta}$ for each $k \in [1,n_\xi]$ (with 1-element width in the $\zeta$ direction, hence the word `pencil').

\item CPU$^{1}$ continues the forward sweep while CPU$^{0}$ starts solving the new tridiagonal system by shifting 1 step from the first block in the $\zeta$, which is the `slide' direction. The `slide' and `pencil' directions can be swapped.

\item CPU$^{1}$ passes data to CPU$^{2}$ for the sliding index $i=1$, receives data from CPU$^{0}$ at the sliding index $i=2$, and continues the forward sweep.
\item The same process is carried out until CPU$^{N_\eta-1}$ reaches the slide index $i=n_\zeta$.
\item CPU$^{N_\eta-1}$ starts the backward sweep at the sliding index $i=n_\zeta$, shares data of size $n_\xi*1$-element containing $\zeta^{N_\eta-1}_1$ for each $k \in [1,n_\xi]$ with CPU$^{N_\eta-2}$, and starts the backward sweep at the sliding index $i=n_\zeta-1$.
\item The backward sweeping process is carried out in the same way as  the forward sweep.
\item Solving the system of tridiagonal matrices is finalized after CPU$^{0}$ finishes the backward sweep at the sliding index $i=1$.
\end{enumerate}  

\subsubsection{Handling shell cut }
\label{sec:shellcut}
\label{sec:shellcut}

%{\color{red} It may be better to include the shell cut figure that you created.}{\color{blue} I made this schematic for visualizing the "shell cut" interface in physical domain. But a depiction like this may lead to many objections. I do not think we should include this. Also, such interfaces are traditionally referred to as "wake cut" in the CFD community, though the term "wake" has no physical significance in our case. - Souvik}

%\begin{figure}
% \centering
% \includegraphics[width=0.8\textwidth]{figures/shell_cut_curv.pdf}
% \caption{Illustration of tridiagonal system solve in the curvilinear domain in $\zeta$ direction with the corresponding CPU ranks.}
% \label{fig:shell_curv}
%\end{figure}

Using the parallel Thomas algorithm with pipelining, we have been able to solve \ref{solve_2}, \ref{solve_3}, and \ref{solve_4} in parallel for $u^{\star}$, $u^{\star \star}$, and $u^{\star \star \star}$. The grid used for the solver before it is rotated about the $x_1$-axis is shown in figure \ref{fig:domtrans}a. The directions parallel and perpendicular to the body surface are denoted by $\zeta$ and $\eta$, respectively. Figure \ref{fig:domtrans}b represents the transformed coordinate, that is obtained using Jacobi transformation. The top and bottom edges of the domain, parallel to $x_1$-axis in figure \ref{fig:domtrans}a, are indicated by the phrase 'Branch cut' (AB and CD). They correspond to the left and right sides of the transformed domain, which are parallel to the $\eta$ direction, indicated by the word 'shell cut'. The body surface seen in the curvilinear domain in figure \ref{fig:domtrans}a is transformed to the top and bottom body surface of the transformed domain \ref{fig:domtrans}b. The words 'shell cut' or 'Branch cut' represent a shared interface among CPUs that cuts through the centerline. A tridiagonal system in the $\zeta$ direction created by discretizing \ref{solve_2} is interrupted by the shell cut on both the left and right sides. To handle the shell cut, the tridiagonal system from one side of the cut is merged with the system on the opposite side. Therefore, the resulting system is twice as large as the system without the cut.\\

\begin{figure}
\begin{minipage}[0.5\linewidth]{0.96\linewidth} % Height = 0.45 & Width = 0.96
\centering
\includegraphics[width=0.5\textwidth,angle=0, clip=true,trim=60mm 40mm 50mm 0mm,scale=1.0]{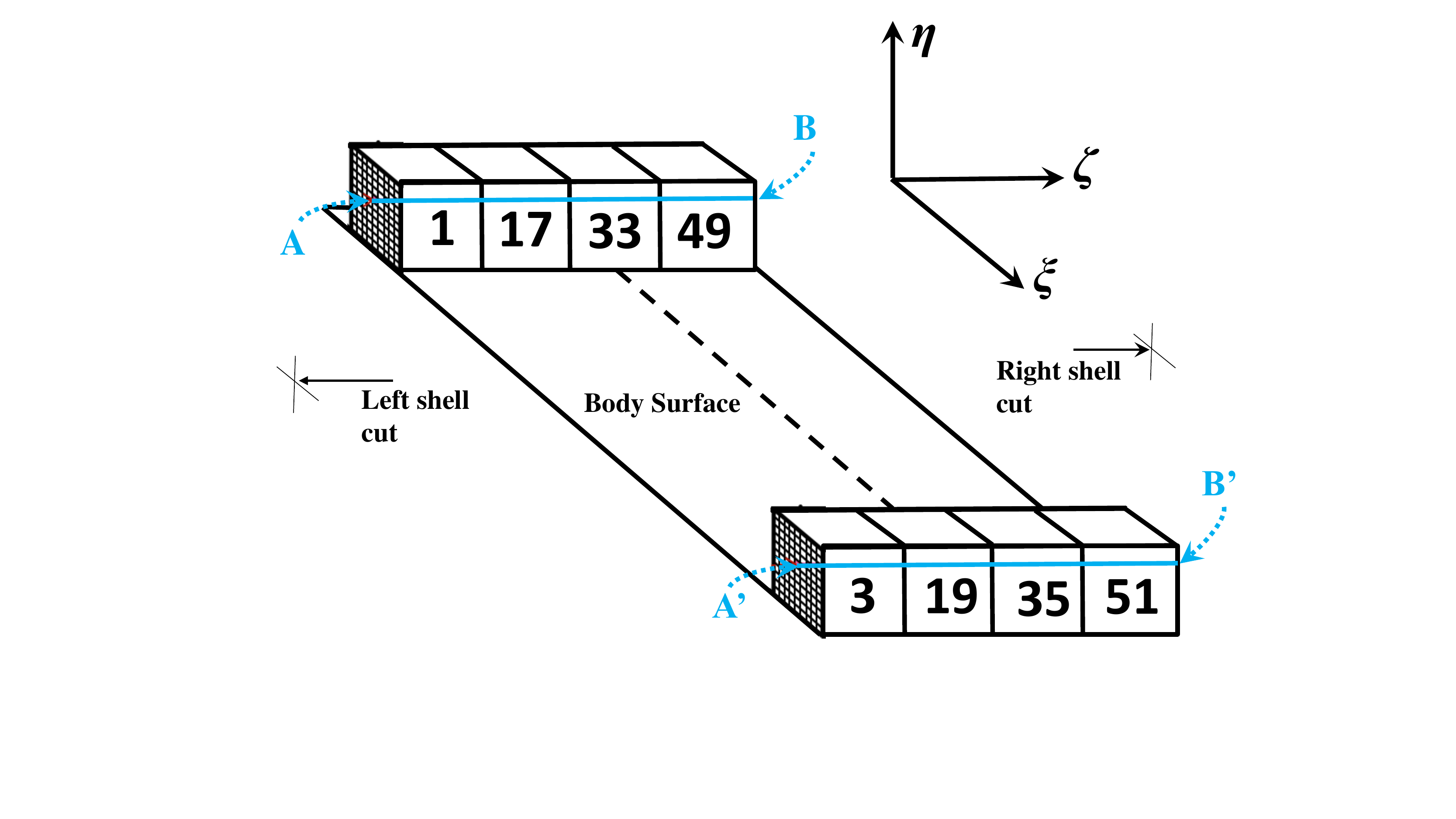}
\end{minipage}
\caption{\label{fig:shellcut} Solving a tridiagonal system across a shell cut. Here (A, A') and (B, B') are the two pairs of surfaces across the left and right shell cuts, respectively, as per the arrangement of processors shown in figure \ref{fig:ThomasParallelled}.}
\end{figure}

For example, consider solving a system in the $\zeta$ direction in figure \ref{fig:shellcut}. In the forward sweeping, CPU$^{1}$ with point A starts solving from this point A and then passes the information to CPU$^{17}$ until the forward sweeping reaches CPU$^{49}$. Then, CPU$^{49}$, with point B passes the information to CPU$^{51}$ that has point B$^{'}$ on the other side of the cut. CPU$^{51}$ keeps performing the forward sweep by sending data to CPU$^{35}$ and so on until the sweep reaches $A^{'}$ owned by CPU$^{3}$. The backward substitution follows the same procedure by starting from point A$^{'}$ and marching until the substitution reaches back to point A. Forward and backward sweeping are done using the pencil-type pipeline Thomas algorithm explained previously.\\

\subsection{Pressure Correction}\label{sec:prcorrect}

The generalized curvilinear solver uses a combination of the ADI-CN-RKW3 methods to obtain $u^{\star \star \star}$, the intermediate velocity. The remaining procedure in the fractional step method is to remove the divergence residual from the projected velocity $u^{\star \star \star}$ at the end of each sub-RKW3 step (denoted as PC in figure \ref{fig:ADI CN RKW3}). This step requires correcting the pressure to account for the divergenceless field. Rewriting equation \ref{eqn:momentum_trans} as \ref{PC_1}; where $\Box$ represents the advection, the diffusion, and the baroclinic terms. \ref{PC_1} is temporally discretized into \ref{PC_2} and \ref{PC_3}. Here, $u^{\star \star \star}_i$ denotes velocity at the third step of the ADI, and ${h}$ is a sub-time step of RKW3.\\

\begin{equation}
  \frac{\partial J^{-1} u_i}{\partial t}  = \Box -\frac{\partial C_{ji} P}{\partial \zeta_j} 
\label{PC_1}
\end{equation}
\begin{equation}
  \frac{J^{-1} u^{\star \star \star}_i - J^{-1} u^{n}_i}{{h}}  = \Box^n -\frac{\partial C_{ji} P^n}{\partial \zeta_j} 
\label{PC_2}
\end{equation}
\begin{equation}
  \frac{J^{-1} u^{n+{h}}_i - J^{-1} u^{n}_i}{{h}}  = \Box^n -\frac{\partial C_{ji} P^{n+{h}}}{\partial \zeta_j} 
\label{PC_3}
\end{equation}
\ref{PC_3}-\ref{PC_2} gives:
\begin{equation}
  \frac{J^{-1} u^{n+{h}}_i - J^{-1} u^{\star \star \star}_i}{{h}}  = -\frac{\partial C_{ji} \delta P^{{h}}}{\partial \zeta_j} 
\label{PC_4}
\end{equation}
\begin{flushleft}
  Taking divergence of \ref{PC_4} gives \ref{PC_5}. Note that $\partial_i u^{n+{h}}_i=0$.
\end{flushleft}
\begin{equation}
  \frac{1}{{h}}\frac{\partial}{\partial \zeta_j} \left[ \frac{\partial \zeta_j}{\partial x_i} J^{-1} u^{\star \star \star}_i   \right] = \frac{\partial}{\partial \zeta_k} \frac{\partial \zeta_k}{\partial x_i} \left[ \frac{\partial}{\partial \zeta_j} J^{-1} \frac{\partial \zeta_j}{\partial x_i} \delta P^{{h}}  \right]
\label{PC_5}
\end{equation}
\begin{flushleft}
  This yields the Poisson equation \ref{PC_6} for pressure correction $\delta P^{{h}}$
\end{flushleft}
\begin{equation}
  \frac{\partial}{\partial \zeta_i}\frac{\partial}{\partial \zeta_j} \left[ G_{ij} {h} \delta P^{{h}} \right] = \frac{\partial}{\partial \zeta_j} \left[ C_{ji} u^{\star \star \star}_i \right]
\label{PC_6}
\end{equation}

Removing divergence from the $u^{\star \star \star}$ field is done by solving \ref{PC_6} for $\delta P^{\bar{h}}$ and computing $u^{n+\bar{h}}_i$ using \ref{PC_7}.
\begin{equation}
  u^{n+\bar{h}}_i = u^{\star \star \star}_i - \frac{1}{J^{-1}}\frac{\partial}{\partial \zeta_j}\left[ C_{ji} \bar{h} P^{\bar{h}} \right]
\label{PC_7}
\end{equation}

The Poisson equation for pressure correction is solved using the Semi-Coarsening Multigrid routine in the HYPRE library \citep{Brown2000}. The divergence-free field $u^{n+\bar{h^{(1)}}}$ marks the end of RKW3 first sub step. We follow the same procedure until $u^{n+\bar{h^{(1)}} + \bar{h^{(2)}} + \bar{h^{(3)}}}$ = $u^{n+1}$ is obtained. Figure~\ref{fig:ADI CN RKW3} illustrates the entire process. HYPRE is a library of scalable linear solvers and multigrid methods \cite{Falgout2000} and \cite{Yang2002}. Generalized curvilinear solver utilizes two solvers provided by HYPRE: 1) SMG, a parallel semi-coarsening multigrid solver for linear systems \cite{Brown2000} and 2) BoomerAMG, a parallel implementation of the algebraic multigrid method \cite{Ruge1987}.\\

\subsection{Simulation details}\label{sec:details}
We have performed simulations for three Rayleigh numbers, $Ra= 10^5,10^7$ and $10^8$, keeping the Prandtl number constant at $Pr=1$.  The radius ratio is also kept to be constant at $\Gamma=0.6$, and an inverse square-law profile of gravity with the radius $g=(r_o/r)^2$ is assumed.  These choices of parameters are aimed to facilitate comparison with  \citet{Gastine2015}. The number of gridpoints used in each direction for different $Ra$ is given in table \ref{tab:results}. The physical curvilinear grid is clustered in the radial direction near the boundaries to resolve the boundary layers near the solid surfaces before performing the Jacobi transformation. The clustering function is given below.
\begin{equation}\label{eqn:stretching}
 r(j)=\frac{\tanh\left[rx_2\left(\frac{j-1}{nx_2}-\frac{1}{2}\right)\right]}{2\tanh\left(\frac{rx_2}{2}\right)},
\end{equation}
here, $rx_2$ is the stretching factor, and $nx_2$ is the number of grid divisions in the radial direction.\\ 

After the transformation, all the grid spacings are unity, and the information about grid stretching is provided effectively through the elongation matrix. At the bottom and top surfaces, a no-slip boundary condition is used ($u_1=u_2=u_3=0$), while the temperatures are fixed at the bottom ($T=1$) and top ($T=0$) surfaces to impose an unstable gradient for maintaining thermal convection. The periodic boundary condition is used for all the variables in the $\xi$ directions. The "shell-cut" boundary conditions are used in the $\zeta$ direction, as explained before in section \ref{sec:shellcut}.  All simulations are started with $u_i=0$ and small random perturbations in the temperature field.\\

We use the following notations for the surface, volume, and time-averaged quantities. 

\begin{equation}\label{eqn:spaceavg}
\left\langle f \right\rangle_s = \frac{1}{4\pi}\int_{0}^{2 \pi}\int_{0}^{\pi} f \;\sin{\theta}\; d \theta\; d \phi,
\end{equation}

\begin{equation}\label{eqn:volumeavg}
\left\langle f \right\rangle = \frac{1}{V}\int_{r_i}^{r_o}\int_{0}^{2 \pi}\int_{0}^{\pi} f \;r^2\; \sin{\theta}\; d \theta\; d \phi\; dr,
\end{equation}

\begin{equation}\label{eqn:timeavg}
\overline{f} = \frac{1}{\tau}\int_{t_0}^{t_0+\tau} f \; d \tau,
\end{equation}
where $V=\frac{4}{3}\pi(r_o^3-r_i^3)$. All the statistical quantities are averaged in time for at least $100$ free fall time ($t_f=d/u_f$) units after the simulation reaches a steady state. Heat transport in the spherical shell is quantified by the Nusselt number $Nu$, which is defined as 
\begin{equation}\label{eqn:nusselt}
    Nu=\frac{\overline{\left\langle u_r T\right\rangle_s}-\frac{1}{\sqrt{RaPr}} \frac{\mathrm{d} \rho}{\mathrm{d} r}}{-\frac{1}{\sqrt{RaPr}} \frac{\mathrm{d} T_c}{\mathrm{dr}}}=-\Gamma \frac{\mathrm{d} \rho}{\mathrm{d} r}\left(r=r_i\right)=-\frac{1}{\Gamma} \frac{\mathrm{d} \rho}{\mathrm{d} r}\left(r=r_o\right),
\end{equation}
where $\rho(r)=\overline{\langle T \rangle_s}$, $\langle \rangle_s$ represents the average over the spherical surface \ref{eqn:spaceavg} and the overbar represents the time average \ref{eqn:timeavg}. Here, $T_{c}$ is the conductive temperature profile for spherical shells with isothermal boundaries given by \ref{eqn:conductive profile}. The thermal conduction equation for a spherical shell with isothermal boundary condition is given by
\begin{equation}\label{eqn:conduction eqn}
\frac{\mathrm{d}}{\mathrm{d} r}\left(r^2 \frac{\mathrm{d} T_c}{\mathrm{~d} r}\right)=0, \quad T_c\left(r = r_i\right)=1, \quad T_c\left(r = r_o\right)=0,
\end{equation}
which yields
\begin{equation}\label{eqn:conductive profile}
T_c(r)=\frac{\Gamma}{(1-\Gamma)^2} \frac{1}{r}-\frac{\Gamma}{1-\Gamma}. 
\end{equation}

\begin{figure}
\begin{subfigure}{0.5\textwidth}
\centering
\includegraphics[width=\textwidth]{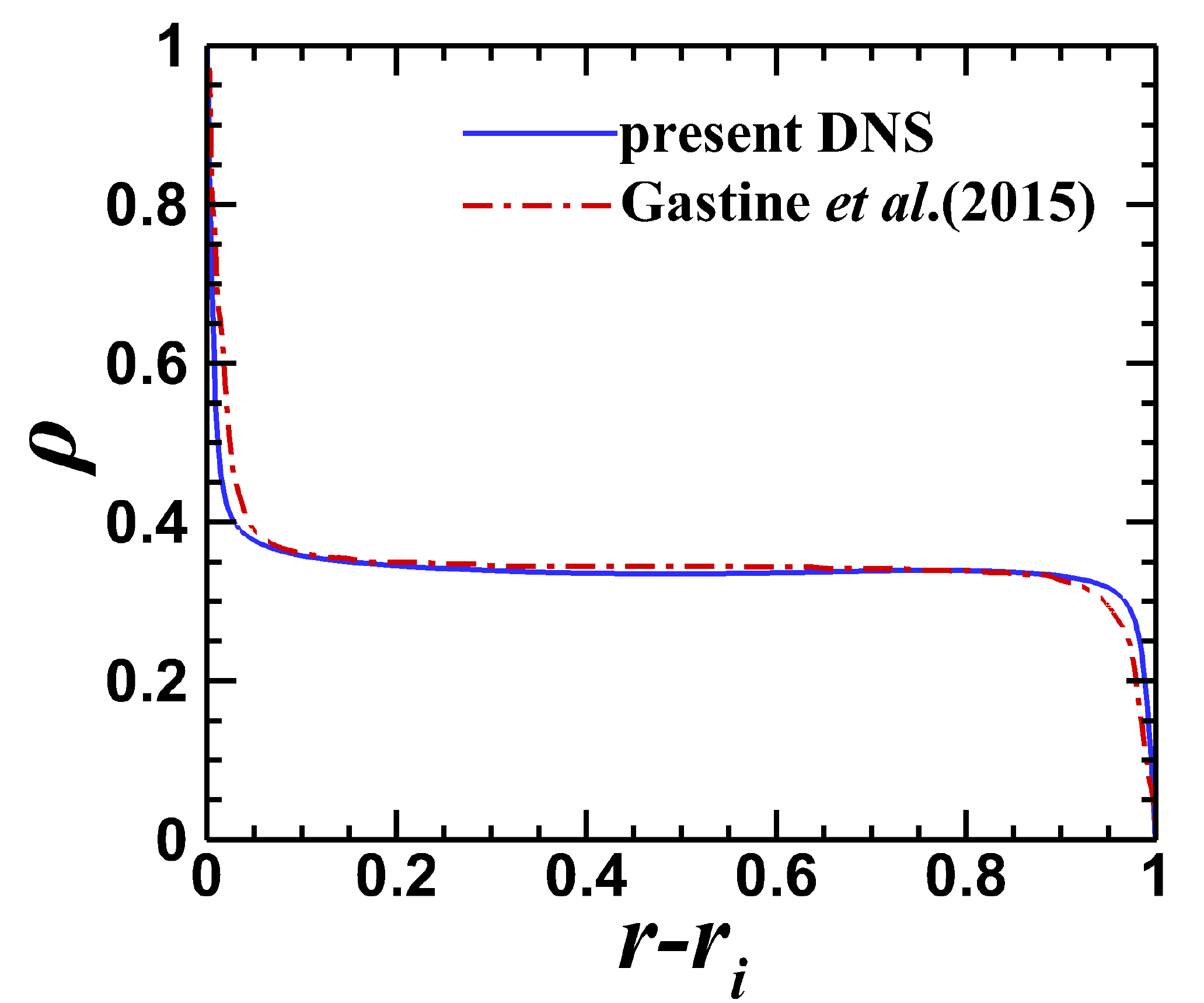}
\caption{\label{fig:T10^7_comparison}}
\end{subfigure}
\begin{subfigure}{0.5\textwidth}
\centering
\includegraphics[width=\textwidth]{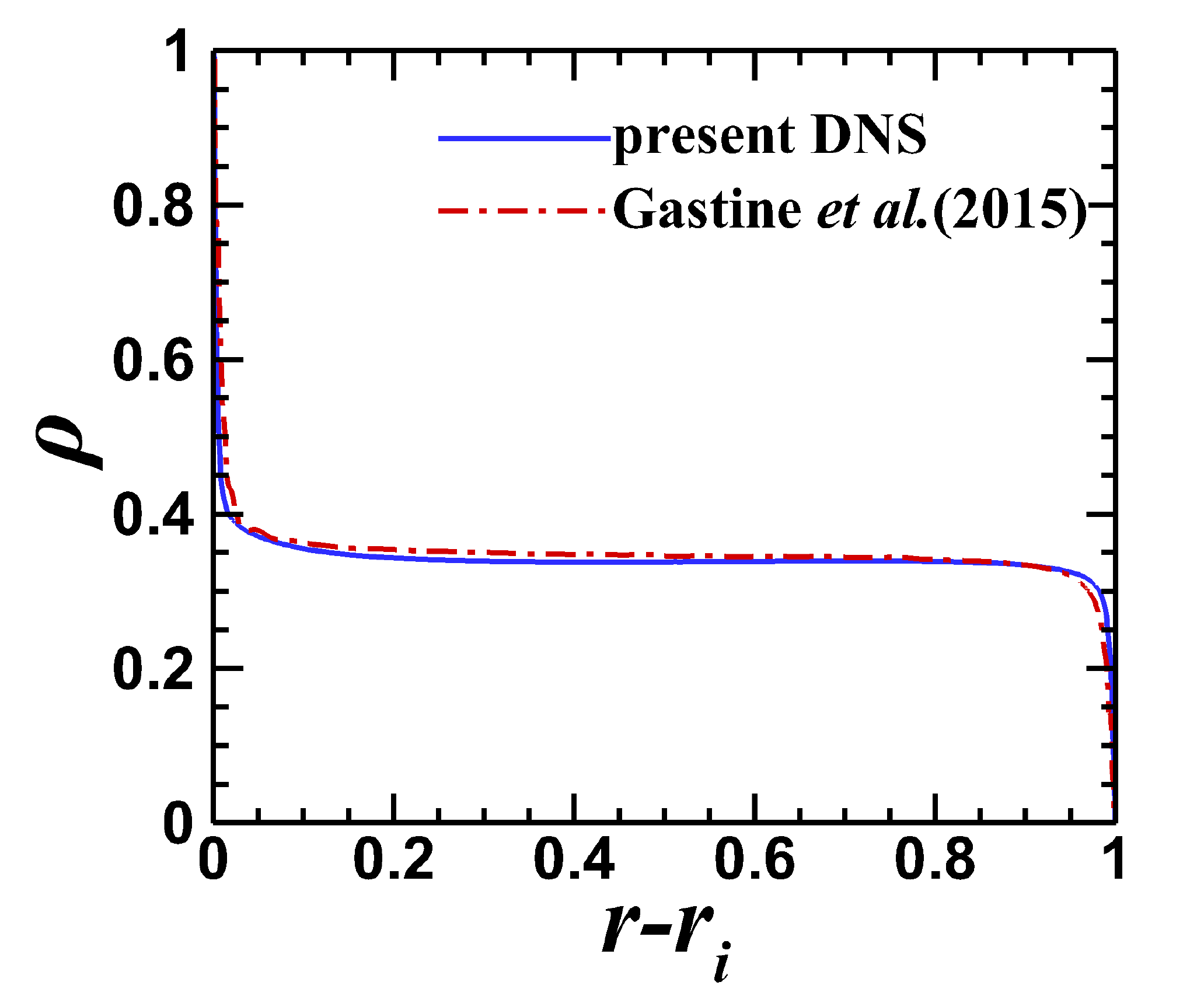}
\caption{\label{fig:T10^8_comparison}}
\end{subfigure}
\caption{Comparison of surface and time-averaged non-dimensional radial temperature profile for the cases at (a) $Ra=10^7$ and (b) $Ra=10^8$.} 
\label{fig:Tcomp}
\end{figure}

\section{Results}\label{sec:results}
This section summarizes the results of the simulations listed in table \ref{tab:results}. We validate our results with those of \cite{Gastine2015}, and further demonstrate the closure of the turbulent kinetic energy budget.\\ 

\subsection{Validation}\label{sec:validation}

\begin{table}[H]
\begin{center}
\begin{tabular}{ c c c c c c}
 \hline
 $Ra$  & Grid & $Nu$ & $Nu$ &$\delta^{T}_{i}/ \delta^{T}_{o}$ & $\delta^{u}_{i}/ \delta^{u}_{o}$ \\
  &($n_{\zeta} \times n_{\eta} \times n_{\xi}$) & (present DNS) & (\citet{Gastine2015}) & &\\
 \hline
 \\
  $10^5$ & $256 \times 128 \times 256$ & 4.60 & 4.71 & 0.095/0.132 & 0.139/0.209\\
  $10^7$ & $512 \times 256 \times 512$ & 17.47 & 17.07 & 0.018/0.031 & 0.076/0.102\\
  $10^8$ & $512 \times 256 \times 512$ & 36.50 & 33.54 & 0.009/0.015& 0.063/0.084\\ 
  \\
 \hline
\end{tabular}
 \caption{Summary table for different $Ra$ along with the grid used for each case. Here, Prandtl number $Pr=1$,  gravity profile, $g=(r_o/r)^{2}$, and radius ratio $\Gamma=0.6$ has been used for all the runs.}
\label{tab:results}
\end{center}
\end{table}
We compare the non-dimensionalized temperature profile variation in the radial direction in figure \ref{fig:Tcomp}. The radial temperature variation is found to be matching with that of \citet{Gastine2015}. Additionally, we compare the $Nu$ obtained from our solver with the values reported by \citet{Gastine2015} at the same $Ra$. For the calculation of $Nu$, we use equation \ref{eqn:nusselt} with $r_i=3$ and $r_o=5$. From the table \ref{tab:results}, it can be observed that $Nu$ matches very well with the values from the \citet{Gastine2015}. \\

\subsection{Flow visualization}\label{sec:visual}
\begin{figure}
\begin{subfigure}{0.5\textwidth}
\centering
\includegraphics[width=\textwidth]{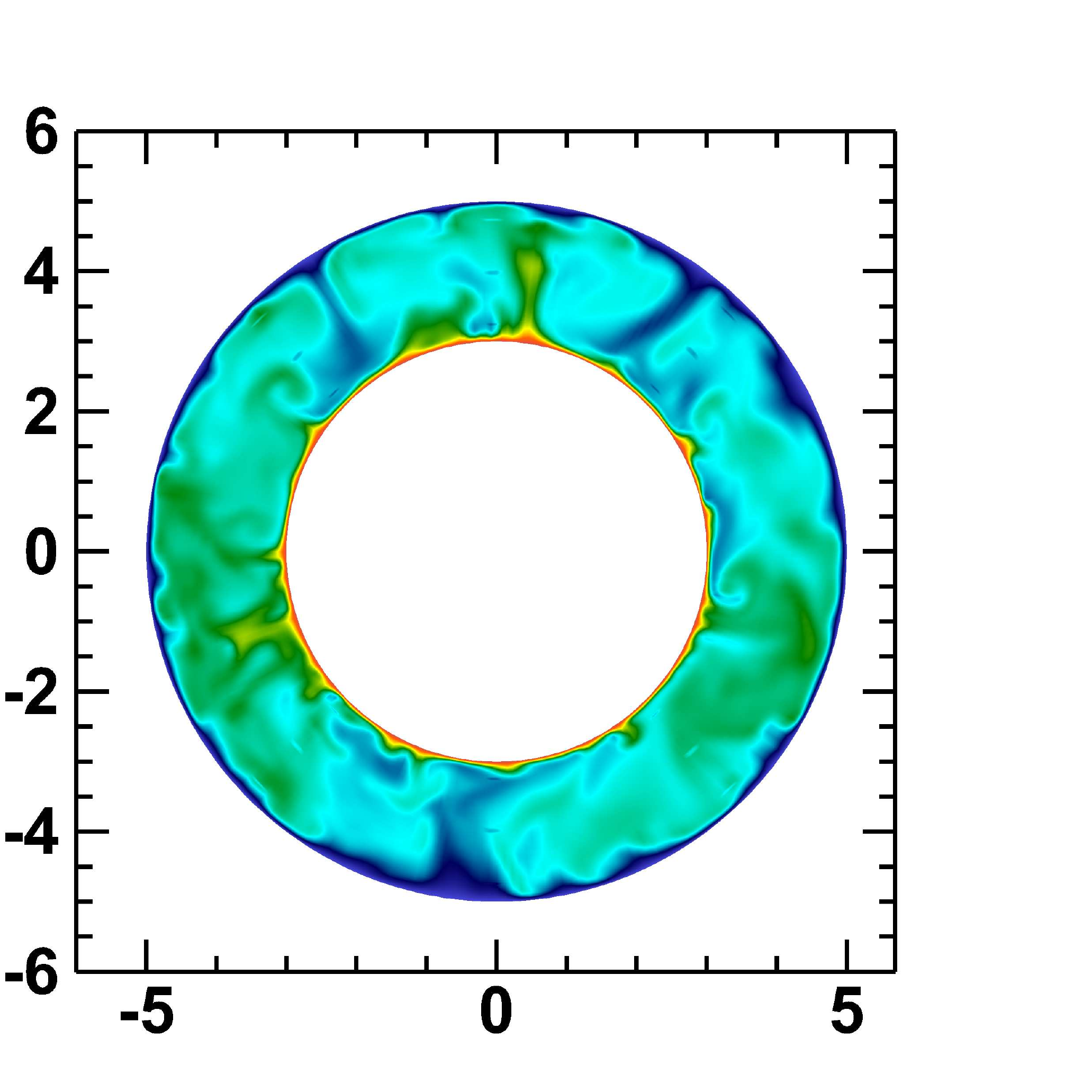}
\caption{\label{fig:10_5_temp_contour}}
\end{subfigure}
\begin{subfigure}{0.5\textwidth}
\centering
\includegraphics[width=\textwidth]{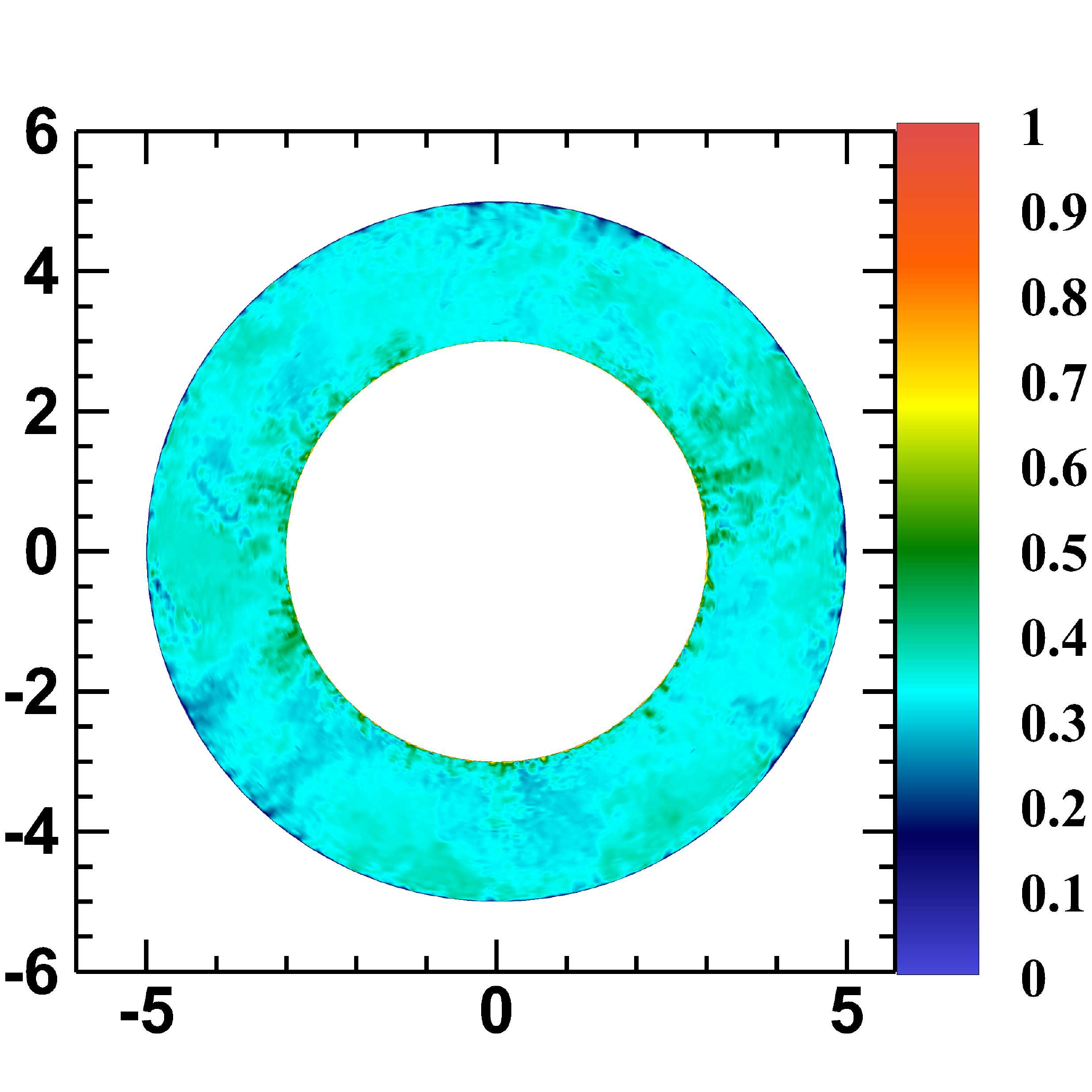}
\caption{\label{fig:10_8_temp_contour}}
\end{subfigure}
\begin{subfigure}{0.5\textwidth}
\centering
\includegraphics[width=\textwidth]{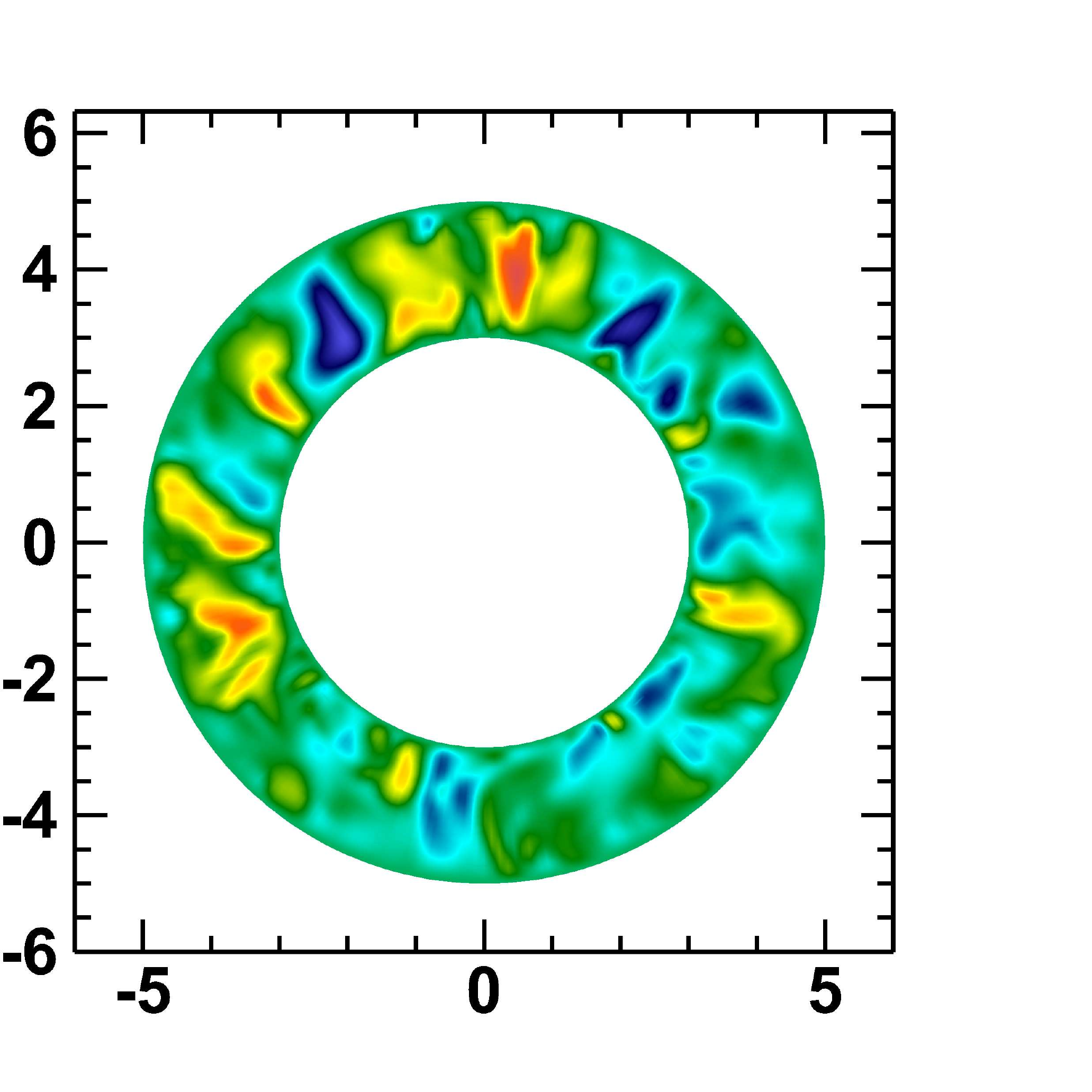}
\caption{\label{fig:10_5_ur_contour}}
\end{subfigure}
\begin{subfigure}{0.5\textwidth}
\centering
\includegraphics[width=\textwidth]{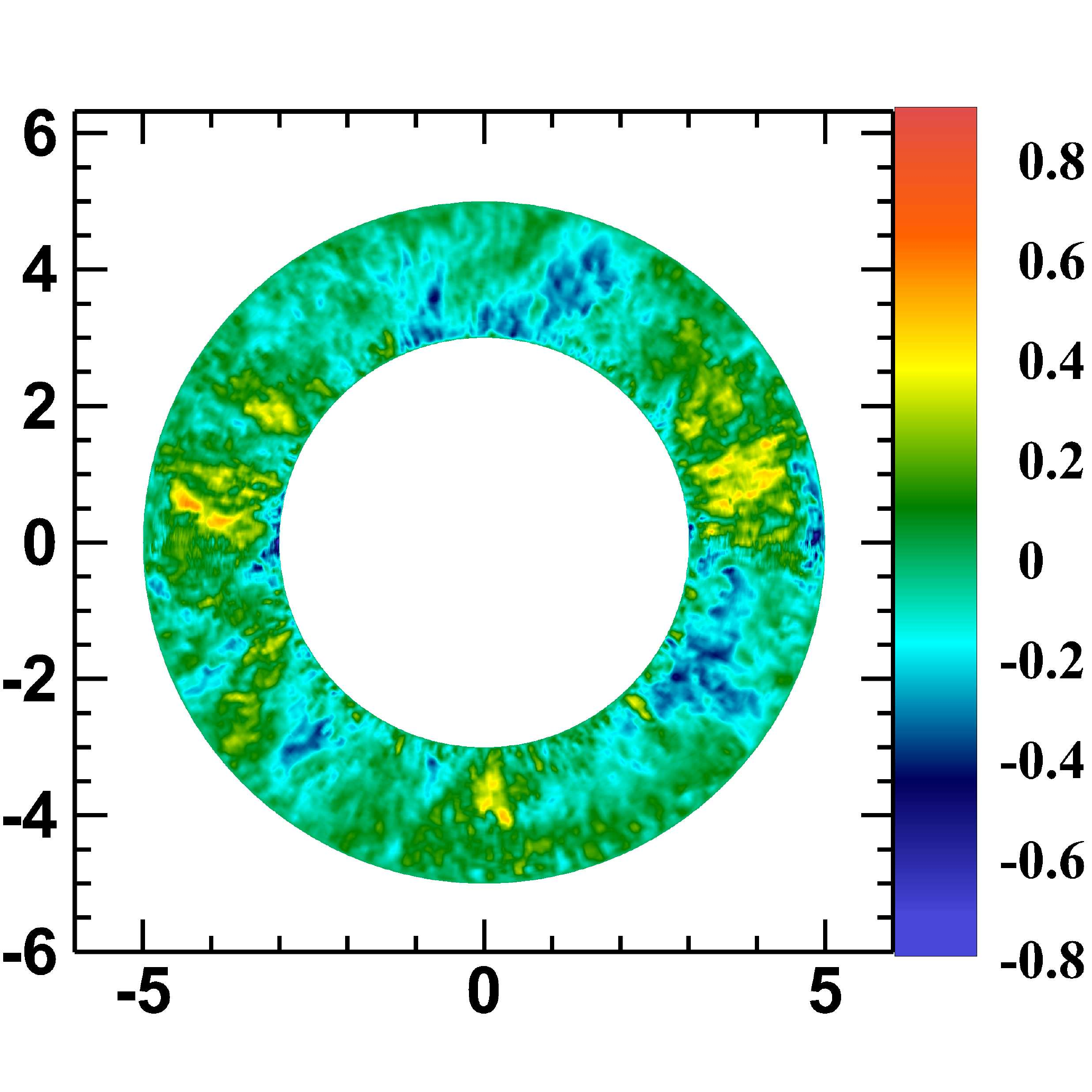}
\caption{\label{fig:10_8_ur_contour}}
\end{subfigure}
\caption{Instantaneous snapshots of  (a,b) the thermal field and (c,d) radial velocity field observed at Rayleigh numbers (a,c) $Ra=10^5$ and (b,d) $Ra=10^8$.}
\label{fig:T_Ur_contour}
\end{figure}

A comparison of the qualitative features of the instantaneous flow and the thermal field with an increase of $Ra$ from $10^5$ to $10^8$ is presented in figure \ref{fig:T_Ur_contour}. For lower thermal forcing at $Ra=10^5$, the plumes generated from the boundaries span the radial extent of the domain, as seen from \ref{fig:10_5_temp_contour}. However, $Ra = 10^{8}$, as shown in \ref{fig:10_8_temp_contour}, the plumes are much smaller with a well-mixed interior. With the increase in $Ra$, turbulence increases, accompanied by higher mixing and generation of smaller scales. The higher $Ra$ case will have a negligible temperature gradient in bulk due to enhanced mixing. In figure \ref{fig:10_5_ur_contour}, the alternating regions with positive and negative radial velocities indicate the presence of structures similar to convective rolls, while figure \ref{fig:10_8_ur_contour} exhibit the presence of small-scale plumes near the boundary at higher $Ra$.\\

\subsection{Boundary layer asymmetry}\label{sec:boundary_layer}

\begin{figure}
\centering
\includegraphics[width=0.5\textwidth]{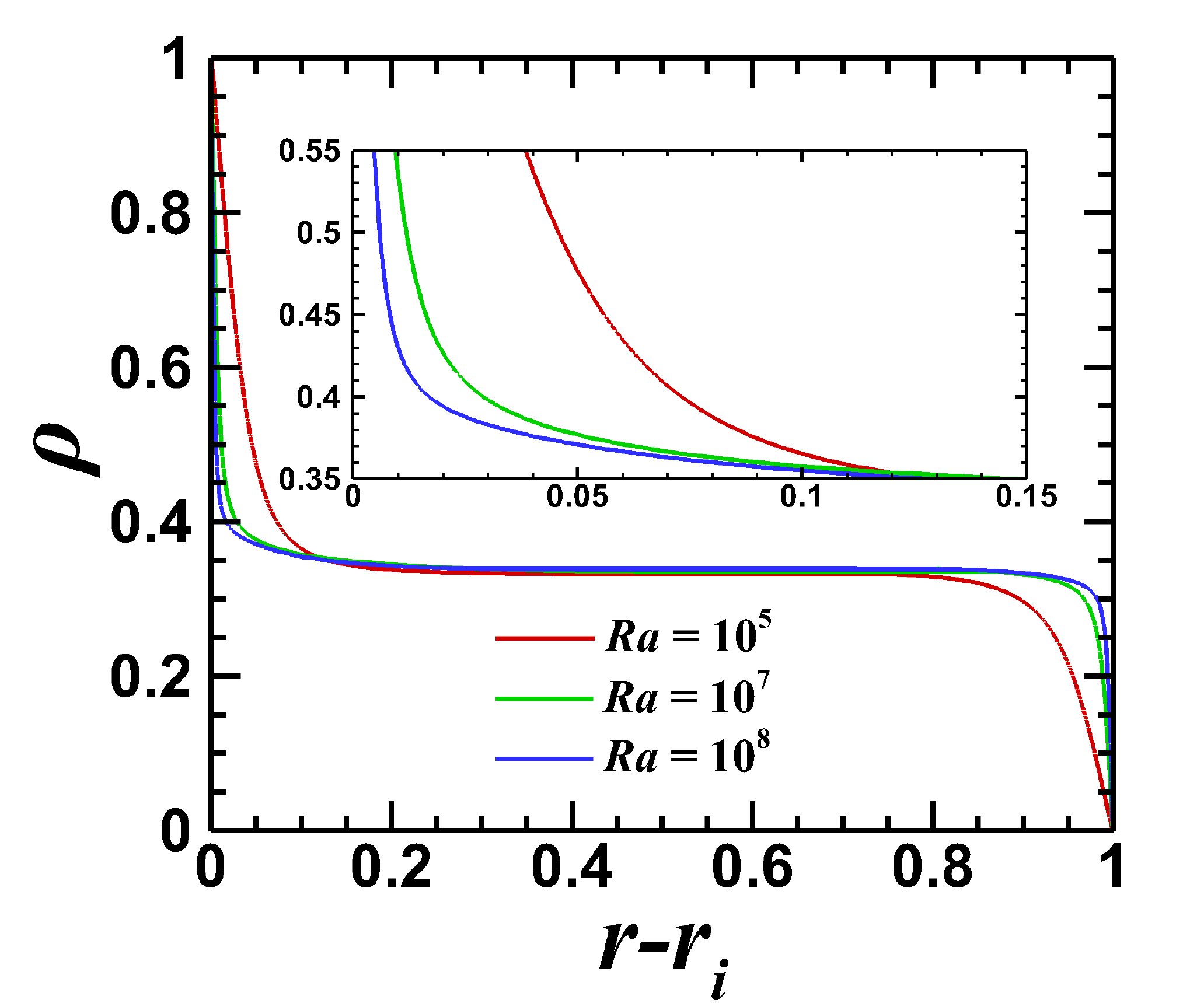}
\caption{Temperature variations observed across different $Ra$. An enlarged view in the inset to demonstrate the increase in gradient with the increase in $Ra$).}
\label{fig:Tempa_all}
\end{figure}

The thermal boundary layer thicknesses ($\delta^{T}_{i}$, inner $\delta^{T}_{o}$, outer) are defined as the distance of the local maximums in the $T_{rms}$ \ref{eqn:Trms} profile from the inner and the outer walls, respectively. The velocity boundary layer thicknesses ($\delta^{u}_{i}$, inner $\delta^{u}_{o}$, outer) are evaluated similarly from the location of the local maximums in the horizontal velocity profile, $u_h$, \ref{eqn:hvelocity} \citep{Long2020}. From the values of boundary layer thickness at the inner and outer boundary from the table \ref{tab:results}, it is visible that there is an asymmetry in the temperature profile at both the inner and outer radius. The total heat flowing in through the inner surface should flow out from the outer surface for thermal equilibrium. In conjunction with the inner spherical shell area being less than the outer spherical shell area, the temperature drop is higher at the inner boundary than at the outer boundary \citealp{Gastine2015}. It is also visible that with an increase in $Ra$, steepening of the temperature profile near the boundaries occurs, as shown in the insets of \ref{fig:Tempa_all}.\\

\begin{equation}\label{eqn:Trms}
    T_{rms}(r)=\sqrt{\left\langle ( T-\langle T \rangle_{s})^2 \right\rangle_s}
\end{equation}
\begin{equation}\label{eqn:hvelocity}
  u_h(r)= \sqrt{\left\langle u_{\theta}^2+u_{\phi}^2 \;\right \rangle_s}
\end{equation}

\subsection{Turbulent kinetic energy budget}
\begin{figure}
\begin{subfigure}{0.5\textwidth}
\centering
\includegraphics[width=\textwidth]{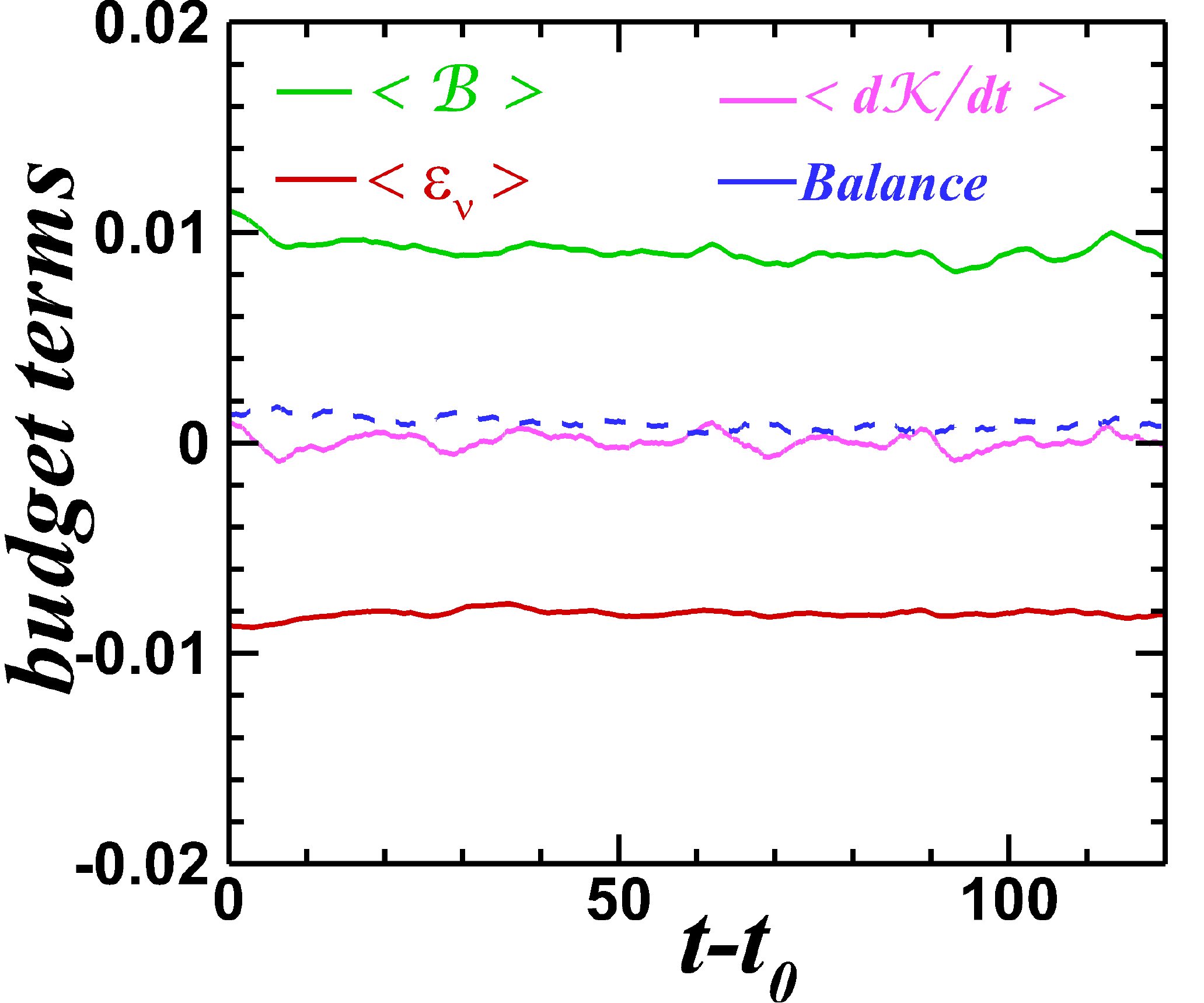}
\caption{\label{fig:budget}}
\end{subfigure}
\hspace{2pt}
\begin{subfigure}{0.5\textwidth}
\centering
\includegraphics[width=\textwidth]{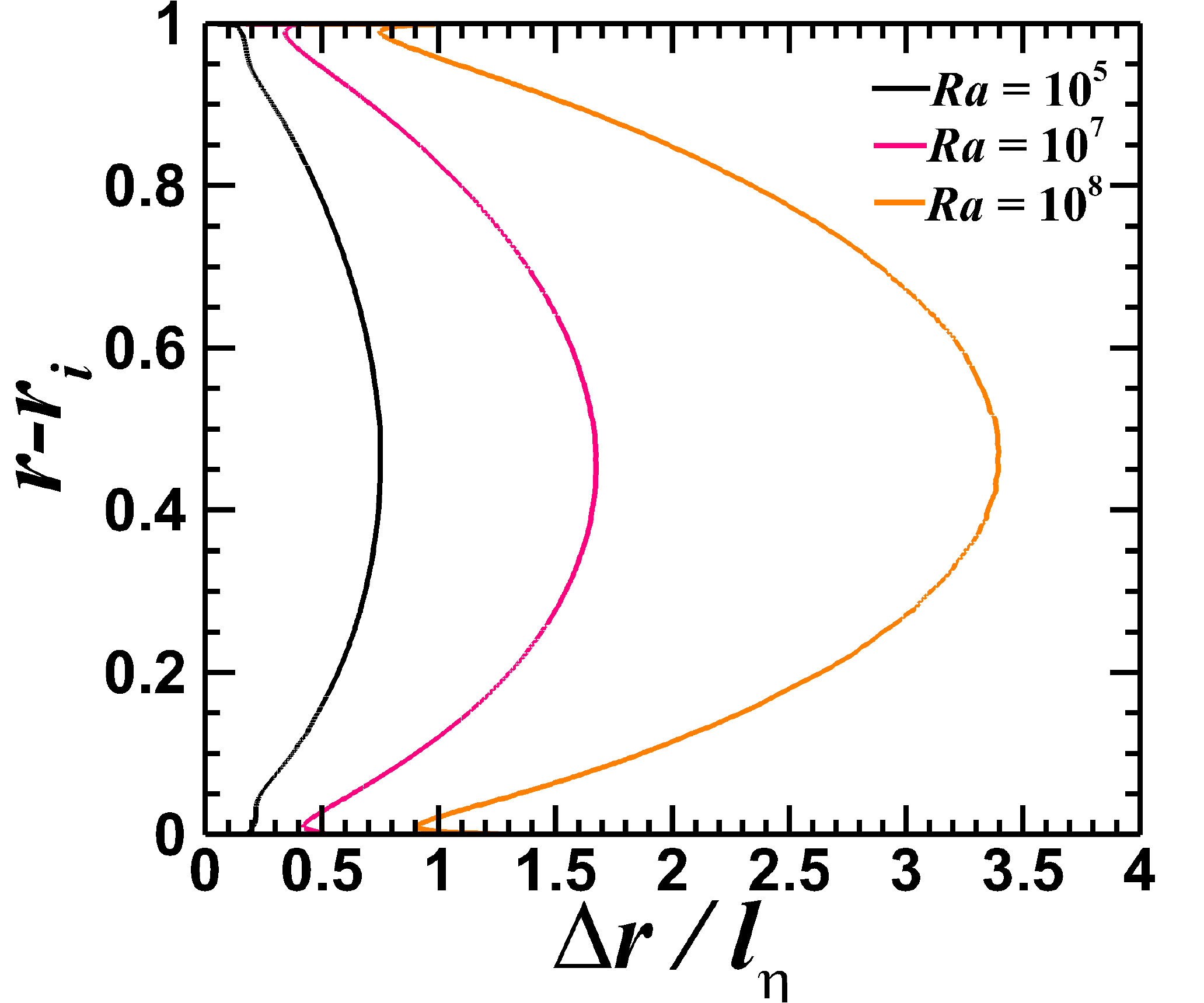}
\caption{\label{fig:dr_l_eta}}
\end{subfigure}
\caption{(\textit{a}) The $t.k.e.$ budget terms for $Ra=10^7$, and (\textit{b}) the radial variation of grid spacing ($\Delta r$) normalized by the Kolmogorov
scale ($l_{\eta}$) as estimated from the spatially averaged dissipation ($\left \langle \epsilon_\nu \right \rangle_s$).}
\label{fig:adding}
\end{figure}

We discuss the turbulent kinetic energy ($t.k.e.$) budget in RBC in this section to review not only the kinetic energy balance but also the adequacy of the resolution of the present simulations. The budget can be expressed as follows: 
\begin{equation}\label{eqn:TKE}
 \left\langle\frac{\mathrm{d\mathcal{K}}}{\mathrm{d} t}\right\rangle  = \left\langle \mathcal{B}\right \rangle-\left\langle \epsilon_\nu \right\rangle,
\end{equation}
where,
\begin{equation}
  \left\langle \mathcal{K} \right\rangle=\left\langle\frac{1}{2}u_iu_i \right\rangle, \; \left\langle \mathcal{B} \right\rangle= \left\langle gu_rT \right\rangle,\; \left\langle \epsilon_\nu \right\rangle =\sqrt{\frac{Pr}{Ra}}\left\langle (\nabla \times u)^2 \right\rangle. 
\end{equation}
In the RHS of equation \ref{eqn:TKE}, the buoyancy flux $\left\langle\mathcal{B} \right \rangle$ is the source term that converts the available potential energy to turbulent kinetic energy to drive the convective motions. This $t.k.e.$ is converted to internal energy by the viscous dissipation term $\left\langle\epsilon_\nu \right \rangle$,\citep{Tennekes1972} which acts as a sink. Buoyancy flux averaged over the whole spherical volume can be expressed as,\\ 
\begin{equation}\label{eqn:bfluxpower}
 \left\langle\mathcal{B} \right \rangle = \frac{4\pi}{V} \int_{r_i}^{r_o} g r^2 \overline{\left\langle u_{r}T\right \rangle_s} \; d r.
\end{equation}
After substituting $Nu$ from \ref{eqn:nusselt} and $T_c$ from \ref{eqn:conductive profile}, we obtain,\\
\begin{equation}\label{eqn:disspower}
 \left\langle\mathcal{B} \right \rangle = \frac{3}{1+\Gamma+\Gamma^2}\frac{1}{\sqrt{RaPr}}(Nu-1)=\sqrt{\frac{Pr}{Ra}} \; \overline{\left\langle(\nabla \times u)^{2}\right\rangle},
\end{equation}
 \begin{equation}\label{eqn:dissratio}
 \chi_{\epsilon_{\nu}}= \frac{\sqrt{\frac{Pr}{Ra}} \; \overline{\left\langle(\nabla \times u)^{2}\right\rangle}}{\frac{3}{1+\Gamma+\Gamma^2}\frac{1}{\sqrt{RaPr}}(Nu-1)}.
 \end{equation}
 The evolution of the $t.k.e.$ budget for the case $Ra=10^7$ is shown in figure \ref{fig:budget}. We evaluate the volume-averaged $t.k.e.$ budget terms when the simulation becomes statistically stationary ($t\geqslant t_0$, where $t_0$ represents the starting time of the simulations). The balance term signifies the difference between the left and right-hand sides of equation \ref{eqn:TKE}. This quantity remains smaller than $5\%$ of $\langle\mathcal{B}\rangle$, indicating sufficient resolution achieved in the simulation to dissipate all the kinetic energy. To further quantify the spatial resolution of the numerical model, the viscous dissipation ratio ($\chi_{\epsilon_{\nu}}$) defined by \ref{eqn:dissratio} is also tested for its closeness to unity. The table \ref{tab:budget} shows that $\chi_{\epsilon_\nu}$ value is near unity for all cases, signifying good spatial resolution. To further test the adequacy of the resolution, the radial grid spacing is compared against the Kolmogorov scale ($l_\eta$) defined by, $l_\eta = \left(Pr/Ra\right)^{\frac{3}{8}}\left(1/\overline{\epsilon_\nu}\right)^{\frac{1}{4}}$. As seen from the figure \ref{fig:dr_l_eta}, the radial grid spacing normalized by $l_\eta$ is near unity for all the $Ra$ cases near the walls, indicating appropriate wall resolution. As seen in figure \ref{fig:dr_l_eta}, the normalized spacing stays below $4$ for all the cases, which is sufficient for accurate calculation of second-order correlations \citep{brucker_2010}. \\
 
\begin{table}[ht]
\begin{center}
\begin{tabular}{ c c c c}
 \hline
 $Ra$ & $\overline{\left\langle\mathcal{B} \right \rangle}$ & $\overline{\left\langle\epsilon_\nu \right \rangle}$ &$\chi_{\epsilon_\nu}$ \\
 \hline
 \\
  $10^5$ & $1.90 \times 10^{-2}$  & $-1.69 \times 10^{-2}$ & 0.97\\
  $10^7$  & $9.25 \times 10^{-3}$  & $-8.17 \times 10^{-3}$ & 1.03\\
  $10^8$  & $6.16 \times 10^{-3}$  & $-5.54 \times 10^{-3}$ & 1.02\\
  \\
 \hline
\end{tabular}
\caption{The turbulent kinetic energy budget terms and the viscous dissipation ratio, $\chi_{\epsilon_\nu}$, at different $Ra$ with $Pr=1$.}
\label{tab:budget}
\end{center}
\end{table}

\section{Conclusion}\label{sec:conclusion}

This present investigation discusses the development of a generalized curvilinear solver for spherical Rayleigh-Bénard convection. Using the Jacobi transformation, the solver transforms a curvilinear domain into a Cartesian domain, and a set of modified governing equations are solved in the Cartesian domain. The solver uses a second-order central differencing scheme for spatial discretization, while for temporal discretization, a combined marching scheme of ADI-CN-RKW3 is used. A parallel Thomas algorithm with pipelining is used for solving the tridiagonal system, which is more efficient and faster as it reduces the idle time for CPUs. In order to remove the divergence residual from the projected velocity in the intermediate field of the fractional step method, Semi-coarsening Multigrid(SMG) routine from the HYPRE library for the pressure correction is used. The solver simulates three Rayleigh number ($Ra$) cases, namely, $10^5$,$10^7$, and $10^8$. The primary emphasis is given to the ability of the solver to predict the heat transfer, quantified by the Nusselt number, $Nu$. Comparing $Nu$ obtained from the numerical simulation with the expected value is not a reliable criterion to assess its validity because even the under-resolved schemes show good closeness with the $Nu$ while producing temperature fields with strong nonphysical oscillations \cite{Kooij2018}. Due to this fact, the solver is not only validated for its $Nu$ but also with the radial temperature profiles from \citet{Gastine2015}. The radial temperature profile and $Nu$ obtained from our solver demonstrate a good match with the results from \citet{Gastine2015}. To further test the spatial resolution, we check on the viscous dissipation ratio's closeness to unity and turbulent kinetic energy budget closure. The $t.k.e.$ budget reveals good closure for all the $Ra$ cases considered. With increased $Ra$, the temperature profile near the boundaries becomes steeper. This is because the fluid near the boundaries is subjected to strong thermal gradients, generating a large buoyancy force that drives the flow. Therefore, the steepening of the temperature profile near the boundaries is evidence of the buoyancy-induced strong convective flow with increased $Ra$. For a particular $Ra$, the temperature profile shows asymmetry due to the difference in area between the spherical inner and outer shell.\\

The majority of the computational methods developed for spherical shell Rayleigh-B\'enard convection employ spherical harmonic decomposition of the solution variables in the angular coordinates $(\theta,\phi)$ while using finite difference or Chebyshev polynomials in the radial direction \citep{busse_1998,christensen_1998,christensen_1999,dormy_1998,glatzmaier_1984,sakuraba_1999,tilgner_1999}. A notable exception to this rule is reported in the work by \citet{kageyama_1995}. However, all these methods were developed for perfectly spherical geometries. The novelty of the present solver lies in its capability to account for the effects of non-spherical geometries in planetary core convection.\\

Our ongoing work is focused primarily on extending the present solver to include the effects of rotation and magnetic field. Future extensions, with further model improvements, should reveal the possible effects of a non-spherical geometry, not only on the convective patterns but also on the self-generated magnetic field  in global numerical dynamo simulations.\\ 

\bibliographystyle{model1-num-names}
\bibliography{refs}

%\section*{Supplementary Material}

%Supplementary material that may be helpful in the review process should
%be prepared and provided as a separate electronic file. That file can
%then be transformed into PDF format and submitted along with the
%manuscript and graphic files to the appropriate editorial office.

\end{document}